\documentclass[twocolumn,numberedappendix]{openjournal}

\usepackage{xcolor}
\usepackage{textgreek}
\usepackage[utf8]{inputenc}
\usepackage[english]{babel}

\usepackage{hyperref}
\hypersetup{
    unicode, 
    colorlinks=true,
    linkcolor=linkcolor,
    citecolor=linkcolor,
    filecolor=linkcolor,
    urlcolor=linkcolor,
}
\usepackage{color,colortbl}
\definecolor{linkcolor}{rgb}{0.0,0.3,0.5}
\usepackage{tensind}
\tensordelimiter{?}
\DeclareGraphicsExtensions{.bmp,.png,.jpg,.pdf}
\usepackage{verbatim}
\usepackage[normalem]{ulem}
\usepackage{orcidlink}
\usepackage{soul}

\usepackage{amsmath}
\usepackage{bm}
\newcommand{\bmtx}[1]{\mathbf{#1}}

\graphicspath{ {./figs/} }

\begin{document}
\title{Blind Deconvolution in Astronomy: How Does a Standalone U-Net Perform?}

\author{Jean-Eric Campagne\orcidlink{0000-0002-1590-6927}}
\email{jean-eric.campagne@ijclab.in2p3.fr}
\affiliation{ Université Paris-Saclay, CNRS/IN2P3, IJCLab,  91405 Orsay, France}

\begin{abstract}
\textbf{Aims:}
This study investigates whether a U-Net architecture can perform \textit{standalone end-to-end blind deconvolution} of astronomical images without any prior knowledge of the Point Spread Function (PSF) or noise characteristics. Our goal is to evaluate its performance against the number of training images, classical Tikhonov deconvolution and to assess its generalization capability under varying seeing conditions and noise levels.

\textbf{Methods:}
Realistic astronomical observations are simulated using the \texttt{GalSim} toolkit, incorporating random transformations, PSF convolution (accounting for both optical and atmospheric effects), and Gaussian white noise. A U-Net model is trained using a Mean Square Error (MSE) loss function on datasets of varying sizes, up to 40,000 images of size $48 \times 48$ from the COSMOS Real Galaxy Dataset. Performance is evaluated using PSNR, SSIM, and cosine similarity metrics, with the latter employed in a two-model framework to assess solution stability.

\textbf{Results:}
The U-Net model demonstrates effectiveness in blind deconvolution, with performance improving consistently as the training dataset size increases, saturating beyond 5,000 images. Cosine similarity analysis reveals convergence between independently trained models, indicating stable solutions. Remarkably, the U-Net outperforms the oracle-like Tikhonov method in challenging conditions (low PSNR/medium SSIM). The model also generalizes well to unseen seeing and noise conditions, although optimal performance is achieved when training parameters include validation conditions. Experiments on synthetic $C^\alpha$ images further support the hypothesis that the U-Net learns a geometry-adaptive harmonic basis, akin to sparse representations observed in denoising tasks. These results align with recent mathematical insights into its adaptive learning capabilities.
\end{abstract}
\begin{keywords}
    {Deconvolution, U-Net}
\end{keywords}

\maketitle

\section{Introduction}
The convolution operation and additive noise corruption are ubiquitous in signal and image processing. In this work, we focus on the latter context, where one observes $\bm{x}_o$ (an $n \times n$ pixel image) as a 2D matrix resulting from the underlying \textit{ground true} signal $\bm{x}_t$, first modified by a convolution operator (or kernel) $\bm{k}$ and then corrupted by additive noise $\bm{n}$. This process can be expressed as:
\begin{equation}
    \bm{x}_o = \bm{k} \ast \bm{x}_t + \bm{n},
    \label{eq:the_pb}
\end{equation}
where $\ast$ denotes the convolution operation. If $\bmtx{K}$ (an $n^2 \times n^2$ matrix) represents the circulant matrix associated with the convolution operator, the above equation can be reformulated using matrix-vector multiplication:
\begin{equation}
    \bm{x}_o = \bmtx{K} \bar{\bm{x}}_t + \bm{n},
\end{equation}
where $\bar{\bm{x}}_t$ is a column vector obtained by stacking the elements of the 2D matrix $\bm{x}_t$. For notational simplicity, we will hereafter use $\bm{x}_t$, without explicitly mentioning the context. Regarding $\bm{n}$, we assume an additive i.i.d. white Gaussian noise, which for instance can model sensor noise during recording.

In astronomical image processing, which is the focus of this work, the convolution kernel---known as the Point Spread Function (PSF)---arises from the optical transfer function of the telescope and atmospheric perturbations for ground-based instruments. The PSF can vary across the image itself for large field-of-view telescopes and may also be wavelength-dependent, which is particularly challenging when dispersing a spectrum in astronomical spectroscopy. The size of the kernel $\bm{k}$ is \textit{a priori} not constrained to match the size of $\bm{x}_o$. For example, motion blur is typically modeled by a small kernel, whereas PSF models are often larger. The range of pixel values in $\bm{x}_o$ depends on the application, e.g., $\mathbb{R}^+$ or $[-1, 1]$.

\begin{figure}[h]
    \centering
    \includegraphics[width=\columnwidth]{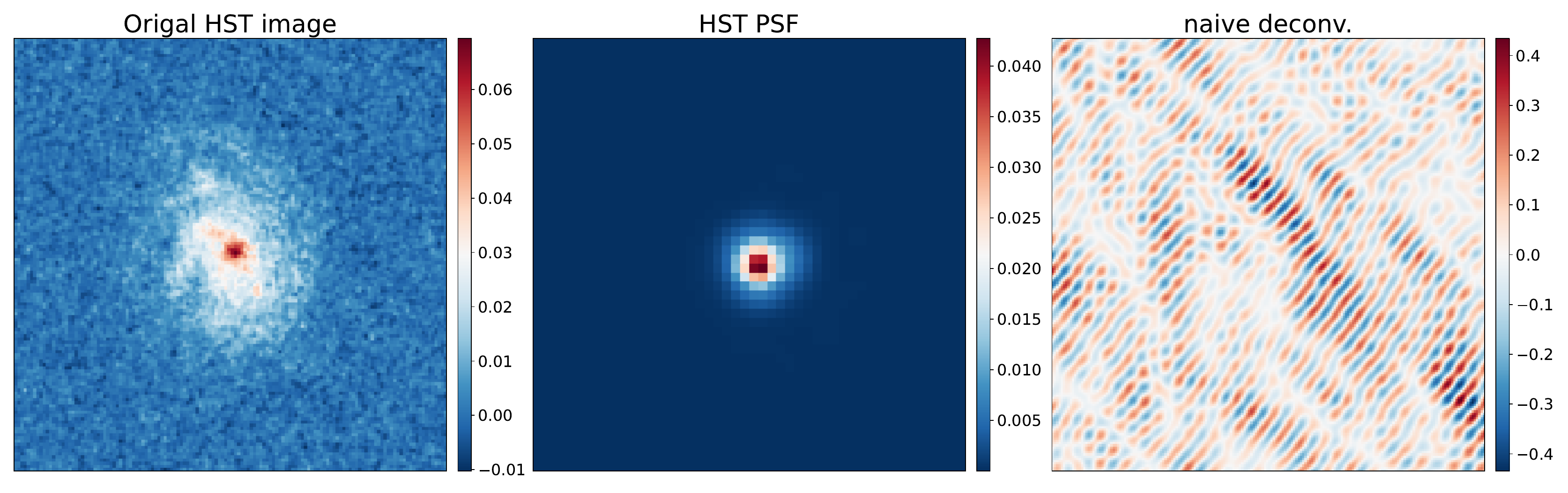}
    \caption{Example of a naive deconvolution of HST image using the associated HST PSF.}
    \label{fig:naive_deconv_example}
\end{figure}

The inverse problem, known as deconvolution, has been addressed across various fields. It has evolved from classical mathematical methods---such as those developed by Norbert Wiener during World War II---to modern deep learning-based approaches. If the kernel is a Dirac delta function, the convolution operation vanishes, and the problem reduces to denoising. When neither the kernel (or PSF) nor the noise characteristics are known \textit{a priori}, the task is referred to as \textit{blind} deconvolution, a well-known ill-posed problem due to the infinite number of possible $(\bm{x}_t, \bm{k})$ pairs that yield the same convolution result: e.g. $(\bm{x}_t, \bm{k})$ and $(\bm{k}\ast\bm{x}_t, \bm{\delta})$ with $\bm{\delta}$ the unit impulse. Even when the kernel is known (\textit{non-blind} case), naive deconvolution is inadequate. Using $\mathcal{F}$ as the Fourier transform, an estimate of the signal, noted $\hat{\bm{x}}$, may be evaluated as\footnote{As a matter of definitiveness, the estimated signal is obtained via $\hat{\bm{x}} = \mathcal{F}^{-1}\{ \mathcal{F}\{\hat{\bm{x}}\}\}$ using the inverse Fourier transform. To avoid heavy notation, we have omitted the arguments of the functions: i.e. in the Fourier space $\mathcal{F}\{\bm{x}\}$ is a function of $\bm{\omega}=(\omega_1,\omega_2)$ for 2D spectrum, and in real space $\bm{x}$ is function of $\bm{u}=(u_1,u_2)$.}:
\begin{align}
    \mathcal{F}\{\hat{\bm{x}}\} = \frac{ \mathcal{F}\{\bm{x}_o\} } {\mathcal{F}\{\bm{k}\}}
    =  \mathcal{F}\{\bm{x}_t\} + \frac{ \mathcal{F}\{\bm{n}\} } {\mathcal{F}\{\bm{k}\}}
    \label{eq:DeconvNaifNonblindFourier}
\end{align}
The second term in the right-hand side of the second equality in Equation~\ref{eq:DeconvNaifNonblindFourier} reveals the dramatic amplification of the noise, overwhelming the signal. This occurs because the Fourier spectrum of the kernel $\mathcal{F}\{\bm{k}\}$ is concentrated at low frequencies (below a certain cut-off frequency), while the white Gaussian noise $\mathcal{F}\{\bm{n}\}$ exhibits a constant power spectral density across all frequencies. Consequently, dividing by $\mathcal{F}\{\bm{k}\}$ disproportionately amplifies the noise at high frequencies, where $\mathcal{F}\{\bm{k}\}$ approaches zero, leading to a reconstructed signal dominated by noise.  An example of this noise enhancement by a naive deconvolution is displayed in Figure~\ref{fig:naive_deconv_example}.
Techniques are therefore required to mitigate these issues both in \textit{non-blind} and \textit{blind} deconvolution.

\subsection{Core Question: U-Net as an End-to-End Blind Deconvolution Network?}
\label{sec:unet_deconv_question}
The U-Net architecture, originally introduced by \cite{ronneberger2015u} for segmentation tasks, has emerged as a powerful tool in image processing, particularly as a \textit{denoiser}. Its mathematical properties have been studied in detail by \cite{kadkhodaie2024generalization}, who characterized what the network learns as a form of \textit{geometry-adaptive harmonic basis}—extending beyond traditional \textit{steerable wavelet bases}. Moreover, in the context of diffusion generative models, where the \textit{denoiser} plays a central role, the authors demonstrated a transition from memorization to generalization as the dataset size increases. These insights naturally lead to a critical question: \textit{Can such a network go beyond denoising and perform \textit{end-to-end blind deconvolution}?}

While U-Net-based networks have been widely adopted as post-processing components in hybrid pipelines (e.g., combined with Tikhonov deconvolution or ADMM, see Section~\ref{sec:relatedworks}), we raise three key questions: (1) Can this architecture perform end-to-end \textit{blind} deconvolution? (2) Does it match the effectiveness of classical methods, such as the Tikhonov functional for which the kernel and noise level are provided? (3) How does dataset size impact its performance,  keeping in mind the memorization-to-generalization transition?

This paper aims to address these questions through numerical experiments, evaluating the U-Net's capacity to generalize and adapt to the challenges of deconvolution. To this end, the remainder of this paper is organized as follows: Section~\ref{sec:math} reviews the classical framework for deconvolution, Section~\ref{sec:relatedworks} discusses related works, Section~\ref{sec:experimental_setup} describes the experimental setup, Section~\ref{sec:results} presents the results, and Section~\ref{sec:summary_discussion} concludes with a summary and discussion.
\section{Backgound in the context of classical framework}
\label{sec:math}
This section provides a brief overview of the mathematical background for deconvolution, focusing on key concepts and classical methods. It is not intended as an exhaustive analysis but rather as a concise introduction to the problem and its regularization techniques. For a more in-depth treatment, readers may refer for insatnce to Chapter~13 of \cite{mallat2009wavelet}.

Starting with the non-\textit{blind} deconvolution inverse problem in a classical analysis framework and using matrix notation, the solution to the naive deconvolution (Equation.~\ref{eq:DeconvNaifNonblindFourier}) can be viewed as the solution to the following least-squares problem:
\begin{equation}
    \hat{\bm{x}}^{\text{naive}} = \arg \min_{\bm{x}} \|\bm{x}_o - \bmtx{K}\bm{x}\|_2^2 = \bmtx{K}^{-1}\bm{x}_o.
\end{equation}
However, as discussed in the introduction, this approach often results in a significant amplification of noise.

To mitigate this issue, regularization techniques—first introduced by \cite{Tikhonov63}—are studied
in classical text books \citep[e.g.,][]{mallat2009wavelet} and in the context of astronomy in the review by \cite{Starck_2002}. These techniques incorporate prior knowledge about the original signal, such as its expected smoothness. The least-squares problem is then reformulated as the minimization of the Tikhonov quadratic functional as
\begin{equation}
    \hat{\bm{x}} = \arg \min_{\bm{x}} \left\{ \frac{1}{2}\|\bm{x}_o - \bmtx{K}\bm{x}\|_2^2 + \lambda \|\bmtx{O}_x \bm{x}\|_2^2 \right\},
    \label{eq:Tikhonov}
\end{equation}
where the first term ensures data fidelity, and the second term acts as a regularizer on $\bm{x}$, with its strength controlled by the parameter $\lambda \in \mathbb{R}^+$. This  yields the following solution\footnote{In matrix notation $\bmtx{A}^\ast$ denotes the conjugate transpose of $\bmtx{A}$.}:
\begin{equation}
    \hat{\bm{x}} = \bmtx{D} \hat{\bm{x}}^{\text{naive}} \quad \text{with} \quad 
    \bmtx{D}  = \left(\bmtx{K}^\ast \bmtx{K} + \lambda \bmtx{O}_x^\ast \bmtx{O}_x\right)^{-1} \bmtx{K}^\ast \bmtx{K}.
    \label{eq:tikhonov_solution}
\end{equation}
The operator $\bmtx{D}$ is recognized as a linear denoiser \citep{mallat2009wavelet}, reducing amplified noise with a regularization strength controlled by the parameter $\lambda$.

In the Fourier domain, this formulation becomes:
\begin{equation}
    \mathcal{F}\{\hat{\bm{x}}\} = \frac{\mathcal{F}\{\bm{k}\}^\ast}{|\mathcal{F}\{\bm{k}\}|^2 + \lambda |\mathcal{F}\{\bm{o}_x\}|^2} \times \mathcal{F}\{\bm{x}_o\},
\end{equation}
where the regularization term $\|\bmtx{O}_x \bm{x}\|_2^2$ is replaced by $\|\bm{o}_x \ast \bm{x}\|_2^2$. This highlights the role of a filter $g$ associated with a spectral window function $W_{o_x}$, such that:
\begin{align}
    \hat{\bm{x}} = \bm{g} \ast \bm{x}_o \quad &\Leftrightarrow \quad \mathcal{F}\{\hat{\bm{x}}\} = \mathcal{F}\{\bm{g}\} \mathcal{F}\{\bm{x}_o\}, \nonumber \\
    &\text{with} \quad \mathcal{F}\{\bm{g}\} := \frac{W_{o_x}}{\mathcal{F}\{\bm{k}\}}.
\end{align}

\cite{Starck_2002} outlines general properties that $W_{o_x}$ must satisfy to serve as an effective regularization window. Common choices for the operator $\bmtx{O}_x$  are high-pass filters such as the gradient and the Laplacian. With such filters, $W_{o_x}$ approaches 1 at low frequencies and 0 at high frequencies, effectively suppressing noise. The hyperparameter $\lambda$ acts as a trade-off between these two regimes: increasing $\lambda$ enhances the regularity of the solution $\hat{x}$.
Alternative window functions are also possible, such as a Gaussian form $W(\omega) = e^{-\alpha \|\omega\|^2}$. 

In the same spirit, the Wiener deconvolution leads to the following spectral window:
\begin{equation}
    W^{\text{Wiener}} = \frac{|\mathcal{F}\{\bm{k}\}|^2}{|\mathcal{F}\{\bm{k}\}|^2 + S\{\bm{n}\}/S\{\bm{x}_t\}},
\end{equation}
where $S\{\bm{n}\}$ and $S\{\bm{x}_t\}$ represent the power spectral densities of the noise and the ground truth signal, respectively. The second term in the denominator corresponds to the inverse signal-to-noise ratio in Fourier space, which tends to zero at low frequencies and infinity at high frequencies. While this satisfies the conditions for a good window, it is more akin to an \textit{oracle}, as it requires \textit{a priori} knowledge of the noise and true image power spectra.

Some limitations of Tikhonov quadratic regularization are highlighted in~\cite{mallat2009wavelet,Sureau2020}. Specifically, when the ground truth signal $\bm{x}_t$ is non-smooth---such as in cases involving rapid contrast transitions---the denoising operator, by suppressing small-scale details, may introduce significant errors. Additionally, the derivative operator $\bmtx{O}_x$ relies on global signal properties, failing to capture local characteristics. 

To address these issues, it has been proposed to replace Equation~\ref{eq:Tikhonov} with the following formulation:
\begin{equation}
    \hat{\bm{x}} = \arg \min_{\bm{x}} \left\{ \frac{1}{2}\|\bm{x}_o - \bm{k}\ast\bm{x}\|_2^2 + \lambda \|\bmtx{o}^s_x\ast \bm{x}\|_1 \right\},
    \label{eq:sparse}
\end{equation}
Here, the operator $\bmtx{o}^s_x$ is carefully chosen to promote a sparse representation of the ground truth signal, as discussed in~\cite{mallat2009wavelet}. This can be achieved through multiresolution analysis using the wavelet transform. Furthermore, the $L^1$-norm in the regularization term is employed to enforce solution sparsity.

\textit{Blind} deconvolution can be seen as an extension of the above problem. For instance, one may use the following generalized objective function:
\begin{equation}
    \mathcal{L}(\bm{x}, \bm{k}) = \ell(\bm{x}_o, \bm{k} \ast \bm{x}) + \lambda_x \, \mathcal{R}_x(\bm{x}) + \lambda_k \, \mathcal{R}_k(\bm{k}),
    \label{eq:blind}
\end{equation}
where $\ell$ is a loss function accounting for the data fidelity term, and $\mathcal{R}_x$ (resp.~$\mathcal{R}_k$) denotes the regularizer for $\bm{x}$ (resp.~$\bm{k}$). \cite{Zhuang2023} discuss some choices for these key elements.

Turning to neural networks \textit{blind} deconvolution of astronomical images  mostly leads to hybrid solutions where the network acts as a denoiser, as mentioned in the following Section~\ref{sec:relatedworks}, while we ask for \textit{end-to-end} realisation. 
\section{Some related works}
\label{sec:relatedworks}
As a preliminary note, the number of articles addressing the deconvolution task may exceed ten thousand over the past decade\footnote{This rough estimate is based on an AI-assisted bibliometric search (e.g., using tools such as \url{https://scispace.com/}).}. This reflects the broad applicability of image deconvolution across diverse fields, including microscopy, medical imaging, photography, video processing, computer vision, archival restoration, and remote sensing. Given the vast scope of the subject, the following discussion does not claim to represent its full diversity. Instead, we categorize methods based on whether they employ neural networks. While our work, presented in Section~\ref{sec:experimental_setup}, relies exclusively on deep networks, we acknowledge the relevance of classical analysis approaches.
\subsection{Non-Neural Network Techniques}
We mention the review by \cite{Satish2020} that explores different methods, notably improvements up on Maximum a Posteriori Estimation for \textit{blind} deconvolution, focusing in the context of non-astronomical images. \cite{Starck_2002} reviewed deconvolution methods applied specifically in astronomy before the deep learning age, with a particular focus on multiresolution wavelet-based transforms. In the exposed classical deconvolution methods, the PSF is assumed to be known, and one of the methods, i.e. Tikhonov regularisation is described in Section~\ref{sec:math}.

More recently, advances have been made to address the challenges of deconvolution in dynamic or complex observational conditions. \cite{Millon_2024} introduced STARRED, a method performing joint deconvolution of all the images in a time series, accounting for epoch-to-epoch variations of the PSF due to atmospheric conditions. The goal is to extract light curves of variable or transient point sources (e.g., stars, supernovae, lensed quasars) while reconstructing the underlying extended scene (galaxies, host galaxies). The method is based on the principles of MCF \citep{Magain_1998} and Firedec \citep{Cantale2016}, paying attention to the Nyquist-Shannon sampling theorem in choosing the PSF modeling.
In parallel, \cite{Berdeu2024} proposed a \textit{blind} deconvolution technique tailored for adaptive optics images. This method estimates both the PSF (including halo) and the sharp object image without prior knowledge, alternating between PSF core estimation, halo modeling, and deconvolution. The result is improved sharpness and the ability to reveal faint companions, all without relying on deep learning but instead exploiting the sharp features of Solar System objects.
\subsection{Neural Network Techniques}
Concerning non-astronomical domains, the reader can benefit from \cite{Singhal2025}'s review, as well as specific works such as \cite{Jin2017} in the context of medical imaging and \cite{Schuler2016, Kupyn2019, Cho2021, Zhuang2023, GUO2025} designed for blur on natural images (e.g. long exposures inducing camera shake) or otherwise degraded images, often assuming simple convolutional kernels and noise models. In contrast, astronomical imaging involves complex, generally spatially-varying PSFs (e.g., due to optics and atmospheric turbulence), making direct application of these methods unsuitable without adapting the physical models and constraints. It is worth noticing that early usages of neural networks can be found for non-\textit{blind} in \cite{Xu2014} and for \textit{blind} deconvolution in \cite{Schuler2016}. \cite{Xu2014} had used a shallow 2-hidden layers network.  \cite{Schuler2016} proceeded to first a feature extraction thanks to a convolutional layer, second a kernel estimation module and finally a image restoration module. These simple architectures reflect the early use of neural networks for deconvolution, which has since evolved significantly in modern works.

\cite{Sureau2020} used a U-Net-based network, to address deconvolution for large surveys with space-variant PSFs. Two strategies were explored: one performs a conventional deconvolution using a closed-form Tikhonov solution, then applies the U-Net as a post-processing refinement (Tikhonet); the other uses an iterative deconvolution framework based on the Alternating Direction Method of Multipliers (ADMM), where the neural network helps in the iterative reconstruction. Both outperformed standard deconvolution but assumed the PSF (or its local variant) was known beforehand. \cite{Nammour2021,Nammour2022} had extended the Tikhonet method using a shape constraint leading to the ShapeNet model. 

Following \cite{Sureau2020}, \cite{Li2023mnras} used a Plug-and-Play ADMM with a known PSF, where a neural network learns denoising priors and hyperparameters from simulated galaxy images. It enables robust deconvolution even with PSF errors but does not perform \textit{blind} deconvolution.

\cite{Akhaury2024} introduced a hybrid deconvolution framework. The process proceeds in three steps: a classical deconvolution (Tikhonov filter) using a known PSF, then denoising refinement with SUNet (a U-Net with Swin Transformer blocks), and a third “debiasing” step based on a multi-resolution support in wavelet space to recover flux/fine structures that may have been lost or biased by the neural network’s output. Tested on HST and VLT images, it efficiently recovers fine structures but does not perform \textit{blind} deconvolution.
\subsection{U-Net Inspired Model as Denoiser}
\begin{table*}
\renewcommand{\arraystretch}{1.5}
\centering
\caption{Use of U-Net variants and their role in selected image deconvolution studies.}
\begin{tabular}{ccp{7cm}}
\textbf{Reference} & \textbf{U-Net Type} & \textbf{Role of U-Net} \\ \hline\hline
\cite{Akhaury2024} & Swin-UNet & Post-deconvolution denoising/correction/restoration, combined with Tikhonov deconvolution and wavelet-based debiasing. \\
\cite{Li2023mnras} & ResUNet & Denoising/regularization inside unrolled ADMM deconvolution \\
\cite{Sureau2020} & Modified U-Net & Post-processing denoiser (Tikhonet) or denoiser-prior in iterative ADMM deconvolution (ADMMnet)\\
\hline\hline
\end{tabular}
\label{tab:unet_deconv_summary}
\end{table*}
It is noticeable that in many works, a modified U-Net is employed, and Table~\ref{tab:unet_deconv_summary} summarizes the use of such networks in the deconvolution of galaxy images. However, none of them use an end-to-end deconvolution network. Instead, the U-Net serves as a \textit{denoiser}, prior, or post-processing component in a hybrid pipeline that combines classical or physics-based deconvolution (or iterative optimization) with deep learning-based correction or regularization. As \cite{Sureau2020} argue, they modified the architecture (called \textit{XDense}) (separable convolutions, dense blocks, average pooling, removal of the global skip connection from input to output) to reduce the total number of parameters, prevent overfitting when data is limited, preserve faint pixel information, improve performance on low signal-to-noise galaxy images, and make the network more suitable for deconvolution (rather than segmentation). \cite{Akhaury2024} based their work on the \textit{SUNet} denoiser by \cite{Fan2022}. The U-Net architecture (encoder/decoder + skip-connections) is globally kept, but convolutional blocks are replaced by transformer-based blocks (Swin Transformer) to better handle deconvolution/restoration tasks. So the inductive bias changes from locality/translation invariance in Convolutional Neural Networks (CNN) to self-attention-based feature extraction with long-range relationships. Changing from CNNs to transformers increases architectural complexity and the number of parameters to train.
\section{Experiment setup}
\label{sec:experimental_setup}
As described in Section~\ref{sec:unet_deconv_question}, we aim to use a U-Net architecture to measure the \textit{standalone} ability of the neural network to perform an \textit{end-to-end blind} deconvolution. For the experimental setup, we have been inspired by \cite{Li2023mnras}, although our perspective is different.
\subsection{U-Net architecture and training}
\label{sec:UNet_training}
The default U-Net architecture used contains 2 encoder and 2 decoder blocks, as well as a single mid-level block \citep{ronneberger2015u}. Each block consists of 2 convolutional layers followed by a ReLU non-linearity and batch normalization. Each encoder block is followed by a $2 \times 2$ \textit{spatial} downsampling and a twofold increase in the number of channels. Each decoder block is followed by a $2 \times 2$ \textit{spatial} upsampling and a twofold reduction of channels. This leads to $1.8\text{m}$ trainable parameters $\bm{\theta}$. This architecture is well-suited for processing images of the sizes considered in the next section. 

The objective function $\mathcal{L}(\bm{\theta})$ used during the training phase of the network $\mathcal{D}_{\bm{\theta}}$ is the Mean Square Error (MSE)\footnote{We also tested a multi-scale version of the MSE loss, as proposed in \cite{Li2023mnras}. However, since the performance obtained with the multi-scale MSE was statistically indistinguishable from that achieved with the standard MSE, we opted to retain the MSE loss. This decision is further supported by \cite{kadkhodaie2024generalization}, which links the training of denoisers to the gradient of the log-likelihood of the noisy image distribution (i.e., the score).}. Our goal is to minimize this loss over a total of $N$ primary images, defined as follows:
\begin{equation}
    \mathcal{L}(\bm{\theta}) = \frac{1}{N}\sum_{i=1}^N \|\bm{x}_t^{i,e} - \mathcal{D}_{\bm{\theta}}(\bm{x}_o^{i,e})\|_2^2,
\end{equation}
where the index $e$ spans the total number of epochs. 

However, rather than using static datasets that are typically prepared once for training, testing, and validation, the term \textit{image forward modelling} (or \textit{simulator}) is more appropriate\footnote{This is an extension of data augmentation, which often uses rotation, translation, resizing, and cropping transformations.}.  For clarity, consider the above expression with $j$ denoting an instance of the couple $(i,e)$ such that
\begin{equation}
\bm{x}_t^j = G_t(\bm{z}^j) \quad \text{and} \quad \bm{x}_o^j = G_k(\bm{z}^{\prime j}) \ast \bm{x}_t^j + G_n(\bm{z}^{\prime\prime j}),
\label{eq:the_generators}
\end{equation}
where, in compact notation, $G_t$, $G_k$, and $G_n$ represent the generators for the new ground truth image, the new PSF, and the noise, respectively, with $(\bm{z}^j, \bm{z}^{\prime j}, \bm{z}^{\prime\prime j})$ being random vectors. This approach is inspired by \cite{Li2023mnras}, but with the key difference that the pairs of simulated ground truth and observed images are \textit{not} fixed prior to the learning phase.

The network implementation and training/testing phases use the \texttt{PyTorch}\footnote{\url{https://pytorch.org/}} framework \citep{PyTorch2019}. All experiments were conducted on a single Nvidia V100-32g GPU at the Jean Zay supercomputer at IDRIS\footnote{\url{http://www.idris.fr/eng/jean-zay/jean-zay-presentation-eng.html}}, a national computing center for the France's National Center for Scientific Research (CNRS). Training was carried out on batches of size 40 for 100 epochs, which was sufficient to achieve loss convergence, using the \texttt{Adam} optimizer \citep{Kingma2015} with the \texttt{ReduceLROnPlateau} learning rate scheduler. The initial learning rate was set to $10^{-2}$, with \texttt{(patience, factor, min\_lr)} parameter values of $(5, 0.5, 10^{-5})$.
\subsection{Image forward modelling}
\label{sec:image_forward_model}
We used the facility offered by the galaxy image simulation toolkit \texttt{GalSim} \citep{ROWE2015}\footnote{\url{https://github.com/GalSim-developers/GalSim}} to read and store information about specific training samples of realistic galaxies, such as the COSMOS Real Galaxy Dataset \citep{Mandelbaum2012_COSMOS}. We preferred real galaxy images over synthetic images, despite the limited dataset sizes. The Hubble Space Telescope (HST) images retained, each with at least $48 \times 48$ pixels, amount to about $54,000$, which makes feasible a study of the performance with respect to the training dataset size. Each HST galaxy image is loaded with its associated HST PSF image and the correlated noise image. After HST PSF normalization, the \texttt{galsim.RealGalaxy} class was used to produce a \texttt{galsim.GSObject}, the key object of \texttt{GalSim}. It is worth mentioning that a naive deconvolution takes place at this stage\footnote{Details are available in the \texttt{GalSim} documentation.}.

The image generator follows a series of steps (see Table~\ref{tab:generator_distrib} in Appendix~\ref{sec:appendix_random_distrib} for the random distributions of the parameters and their default values) before convolving the result with a new PSF.
First, random transformations—rotation, translation, shear, and magnification—are applied to the image.
The transformed image is then projected into an array of size $\texttt{fov} \times \texttt{upsampling}$, where $\texttt{fov}$ represents the field of view of the new image in pixels (default: $48$) and $\texttt{upsampling}$ is the upsampling factor (default: $4$).
Finally, the enlarged image is convolved using the HST PSF, simulating an observation by the Hubble Space Telescope\footnote{See the note above regarding the default naive deconvolution approach.}.

Unlike \cite{Li2023mnras}, we chose to use the HST pixel scale of $0.03$ arcsec instead of an LSST-inspired value, as the \texttt{GalSim} drawing function would otherwise result in a zero-padded image with only a small fraction of non-null pixel values\footnote{This is also why we only used rotation angles as integer multiples of $\pi/2$, equivalent to horizontal or vertical flips.}. Finally, the image pixel values are adjusted to the $[-1, 1]$ range. Although this is not realistic for real astronomical images, it matches the Gaussian white noise setup used in our tests.

The simulator then generates a new PSF, accounting for both optical and atmospheric aspects. For optical characteristics, using the \texttt{galsim.OpticalPSF} function, the generator randomly samples the following parameters:~(1)~\textit{defocusing}, to simulate deviations from the ideal focal plane;~(2)~\textit{optical aberrations}, including spherical aberration, biaxial astigmatism, coma, and trefoil, to model common imperfections in optical systems; and~(3)~\textit{obturation ratio} and~$\lambda/D$, which determine the diffraction-limited resolution and the overall PSF shape.
An example is shown in Figure~\ref{fig:psf_optics_exemple_single} (more examples are provided in Appendix~\ref{sec:aap_optical_PSF}).
\begin{figure}[h]
    \centering
    \includegraphics[width=0.7\columnwidth]{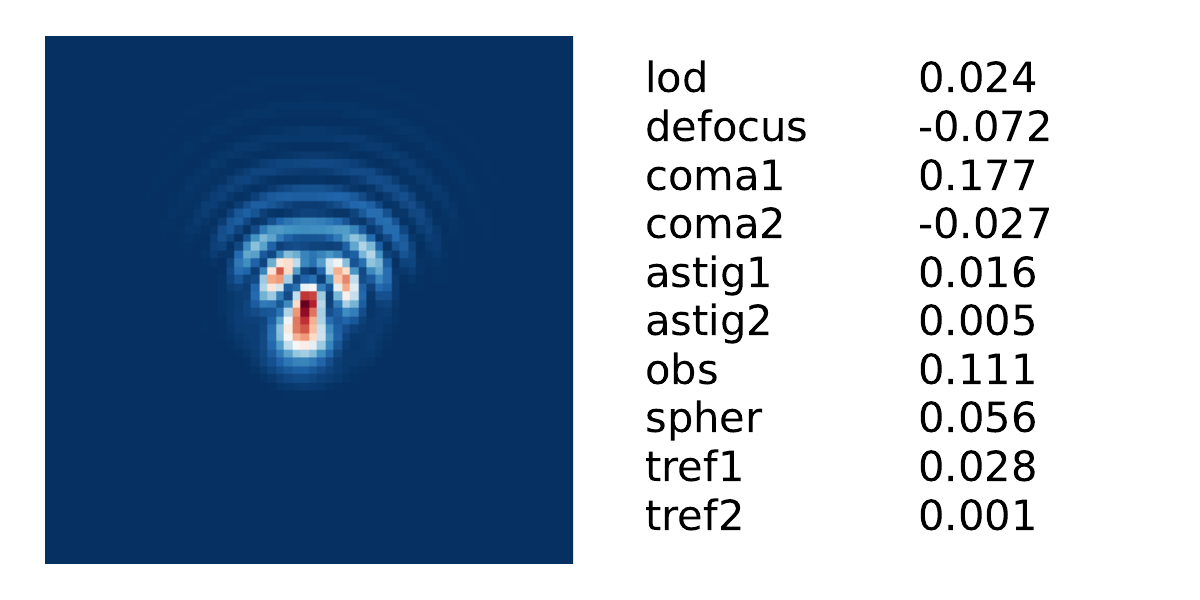}
    \caption{Example of an optical PSF (zoom $\times 3$) obtained from \texttt{galsim.OpticalPSF}.}
    \label{fig:psf_optics_exemple_single}
\end{figure}

\begin{figure}[h]
    \centering
    \includegraphics[width=0.99\columnwidth]{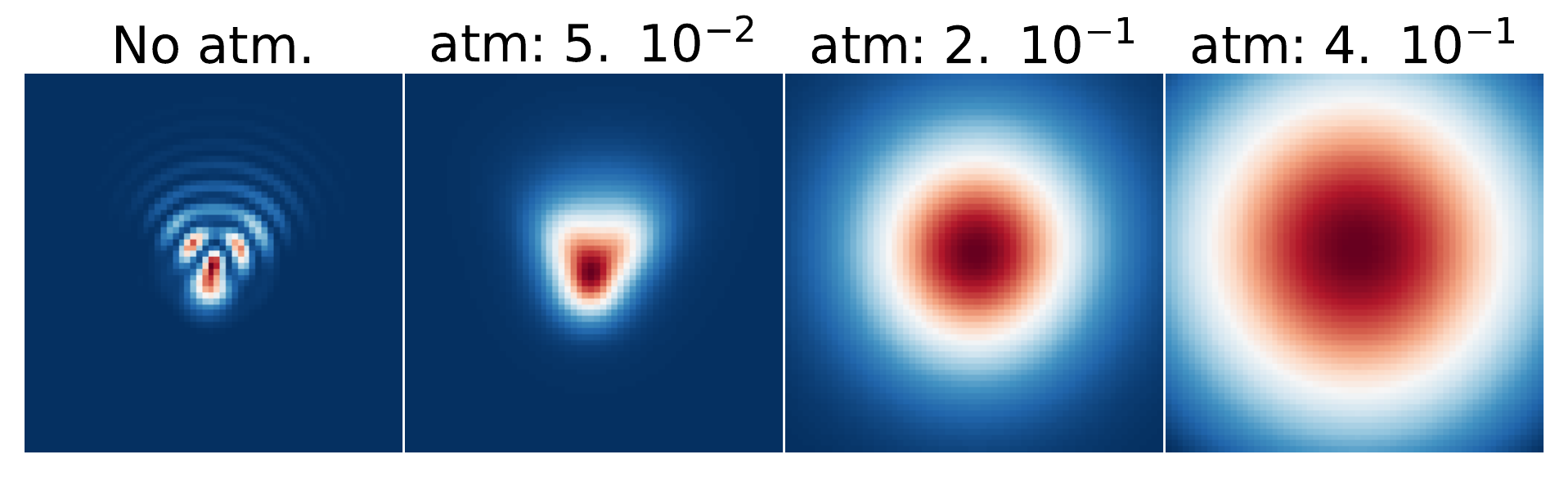}
    \caption{Examples of effects of atmospheric seeing (the FWHM of Kolmogorov function is noted \texttt{atm}) on the optical PSF shown on Figure~\ref{fig:psf_atm_exemple}.}
    \label{fig:psf_atm_exemple}
\end{figure}

For atmospheric perturbations, we employ the Kolmogorov surface brightness profile with a randomly sampled seeing (FWHM) value~(\texttt{galsim.Kolmogorov}). A random shear is also applied to simulate atmospheric variations that introduce shape noise. The final PSF is obtained by convolving both optical and atmospheric contributions. Examples are shown in Figure~\ref{fig:psf_atm_exemple}.
This methodology ensures that the generated PSFs reflect a diverse range of optical and atmospheric conditions, thereby enhancing the robustness of the deconvolution process under study. 

The final steps of the generator apply non-overlapping average pooling with a $4 \times 4$ kernel (matching the $\texttt{upsampling}=4$ factor cited above), reducing the image to a spatial resolution of $48 \times 48$ pixels (default). Subsequently, additive i.i.d.\ Gaussian white noise is injected with a standard deviation uniformly sampled from the interval $[0, 1]$. This noise generation method follows \cite{Campagne_2025} and contrasts with \cite{Li2023mnras}, who use a fixed standard deviation inspired by LSST electronic readout noise\footnote{LSST stands for Legacy Survey of Space and Time, although this acronym is more commonly used to refer to the survey performed by the Simonyi Survey Telescope at the Vera C. Rubin Observatory, equipped with the LSST Camera. See \url{https://rubinobservatory.org/} for more information.}.
\section{Experiment results}
\label{sec:results}
In the following, we present results obtained after training U-Net models as described in Section~\ref{sec:UNet_training}. All figures are generated from Python notebooks available in the companion GitHub repository (Section~\ref{sec:data_availability}). As a reminder, unless otherwise specified, the default maximal values for the Full Width at Half Maximum (FWHM) of the Kolmogorov function (representing the seeing) and the standard deviation of the Gaussian white noise are set to $0.4$ and $1.0$, respectively.
The default training dataset consists of $40,000$ images, selected from a total of approximately $54,000$ available images. An additional $1,000$ images, generated using the same forward modelling process described in Section~\ref{sec:image_forward_model}, are reserved exclusively for the validation dataset. Importantly, these validation images were never used during any model training phase.
\subsection{Galaxy image U-Net deconvolutions: some illustrations}
\label{sec:deconv_some_illustrations}
\begin{figure*}
    \centering
    \includegraphics[width=0.75\linewidth]{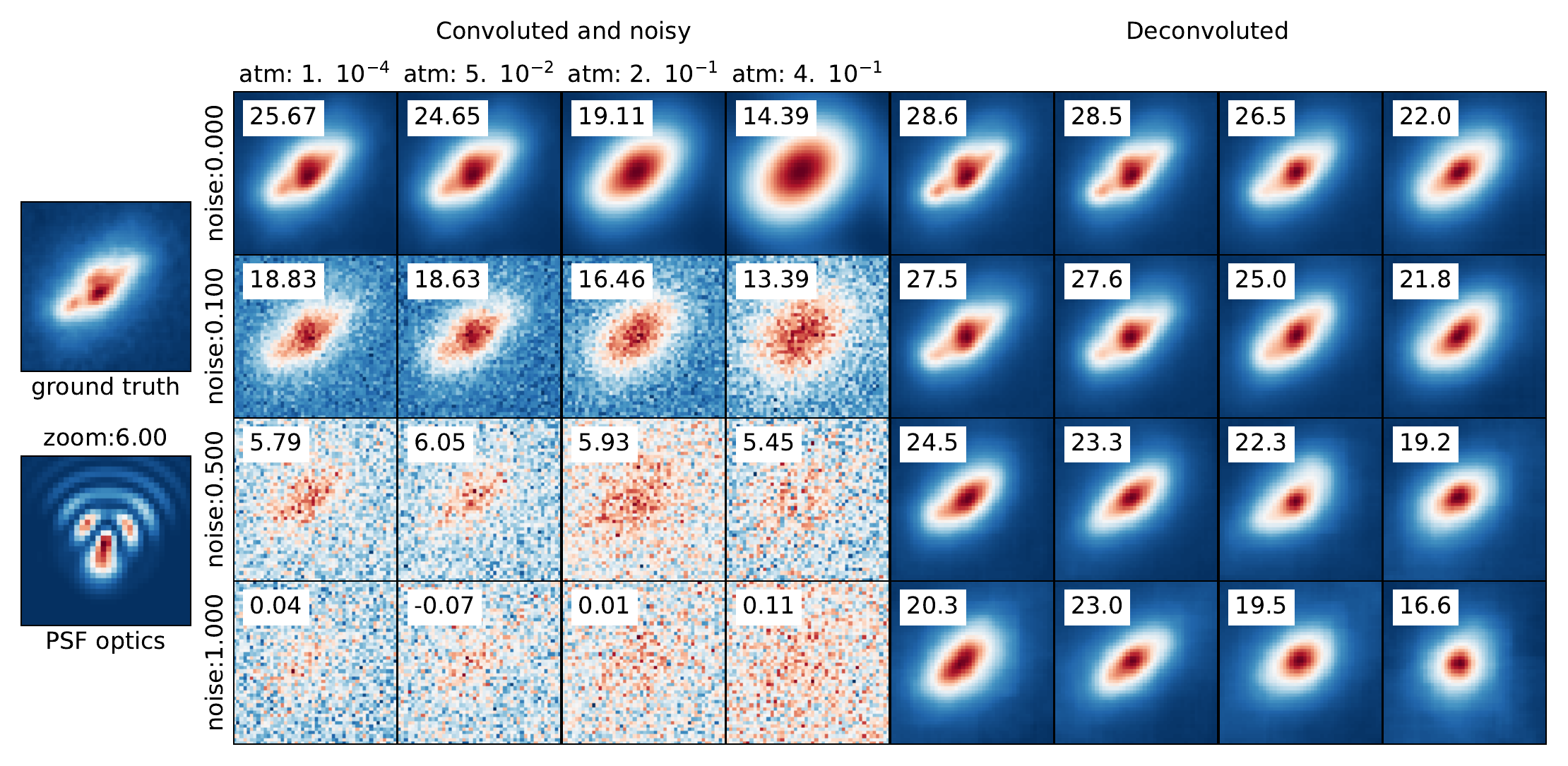}
    \caption{Examples of U-Net deconvolution (see text).}
    \label{fig:exemple1_deonv}
\end{figure*}

After setting a series of parameters---first to define the optical PSF properties and second to modify the original HST galaxy image---we can then simulate varying seeing and noise conditions. This process enables us to generate a set of observed images from a ground truth reference, which are subsequently processed using the same U-Net model for deconvolution. Figure~\ref{fig:exemple1_deonv} provides illustrative examples of this workflow.

The two small images on the left show the ground truth ($48 \times 48$ pixels) and a $\times 6$ zoom of a $32 \times 32$ pixel patch extracted from the optical PSF, which is $192 \times 192$ pixels. The $4 \times 8$ grid of $48 \times 48$ pixel images is divided into two $4 \times 4$ sub-grids:
\begin{itemize}
    \item {Left grid} (\textit{``Convoluted and noisy''}): Each image corresponds to a specific combination of seeing conditions (atmospheric perturbation factor, \textit{``atm''}, ranging from $10^{-4}$---no perturbation---to $0.4$, the maximum value used during U-Net training) and noise levels (standard deviation, \textit{``noise''}, ranging from $0$---no added noise---to $1.0$, the maximum value used during training).
    \item {Right grid} (\textit{``Deconvoluted''}): The corresponding deconvoluted images, processed by the U-Net model.
\end{itemize}
For each image (\texttt{img}) in both grids, we computed the Peak Signal-to-Noise Ratio (PSNR) using the Mean Squared Error (MSE) metric, with the ground truth image (\texttt{ref}) as reference\footnote{As a reminder, the pixel values of the ground truth image are in the range $[-1, 1]$.}:
\begin{equation}
\text{PSNR}(\texttt{img}, \texttt{ref}) = -10 \log_{10} \left( \text{MSE}(\texttt{img}, \texttt{ref}) \right).
\end{equation}
While the PSNR is typically used for noisy images---where it simplifies to $-20 \log_{10} \sigma$ for additive Gaussian noise with standard deviation $\sigma$---we adopt the PSNR notation here as a convenient expression of the MSE in decibels (dB).

Each pair of images---one from the left grid (\textit{``Convoluted and noisy''}) and its corresponding deconvoluted image from the right grid---can be directly compared, both visually and through the quoted PSNR values, to assess the model's performance under varying degradation conditions.
One can observe the following:

\begin{itemize}
    \item {Left grid} (\textit{``Convoluted and noisy''}):
    The images become progressively blurrier and noisier as the \texttt{atm} and \texttt{noise} parameters increase.
    The PSNR values, shown at the top of each image, decrease with increasing noise and atmospheric perturbations, as expected due to the degradation of image quality caused by noise and convolution.

    \item Right grid (\textit{``Deconvoluted''}):
    The deconvoluted images exhibit significant improvement compared to the noisy images.
    Their PSNR values are {substantially higher} than those of their corresponding convoluted and noisy counterparts, clearly indicating a superior reconstruction quality.
    Even under high noise and perturbation conditions, the U-Net model successfully restores images, although the quality slightly degrades with increasing perturbations.
\end{itemize}
These first results indicate the ability of the U-Net model to deconvolve galaxy images degraded by noise and atmospheric perturbations. The visual results and PSNR values show that the model is robust even under challenging conditions, although the quality of the deconvolution slightly decreases with increasing perturbations and noise.

In the following, we extend our analysis beyond these initial results, which were obtained using a single ground truth image and evaluated through visual inspection and PSNR metrics. 
\subsection{Performance evaluation using PSNR and SSIM metrics: impact of training set size}
\label{sec:PSNR_SSIM_vs_Ntrain_perf}
\begin{figure*}
    \centering
    \includegraphics[width=0.35\linewidth]{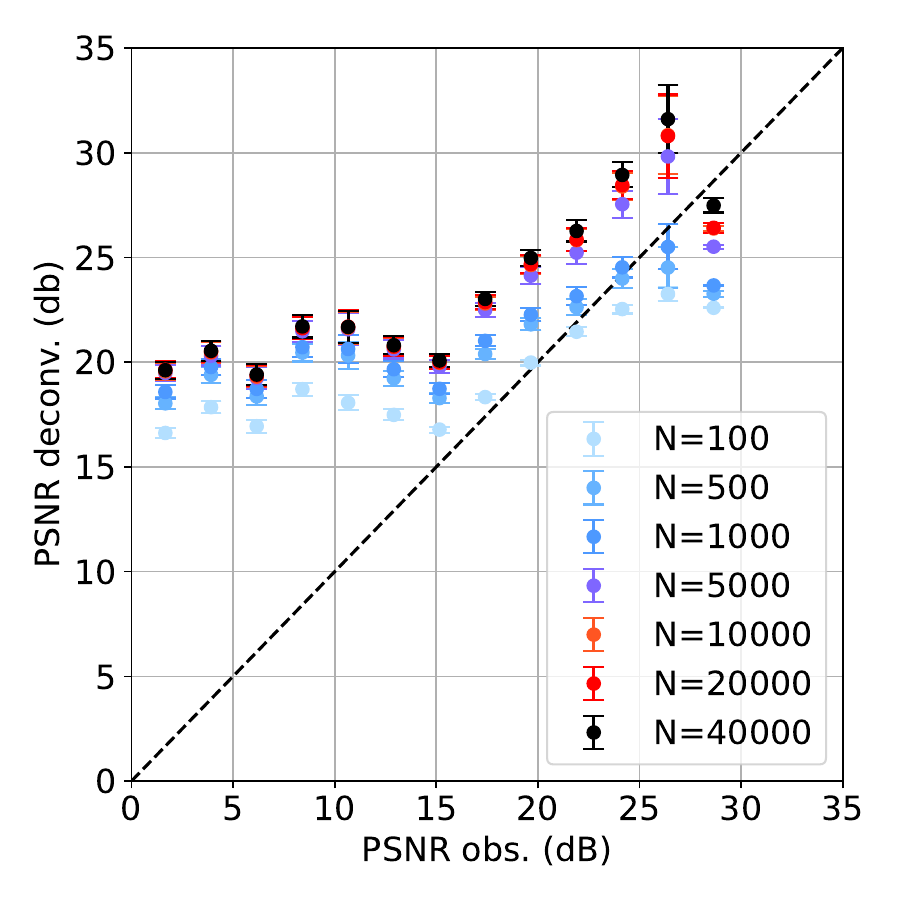}
    \includegraphics[width=0.35\linewidth]{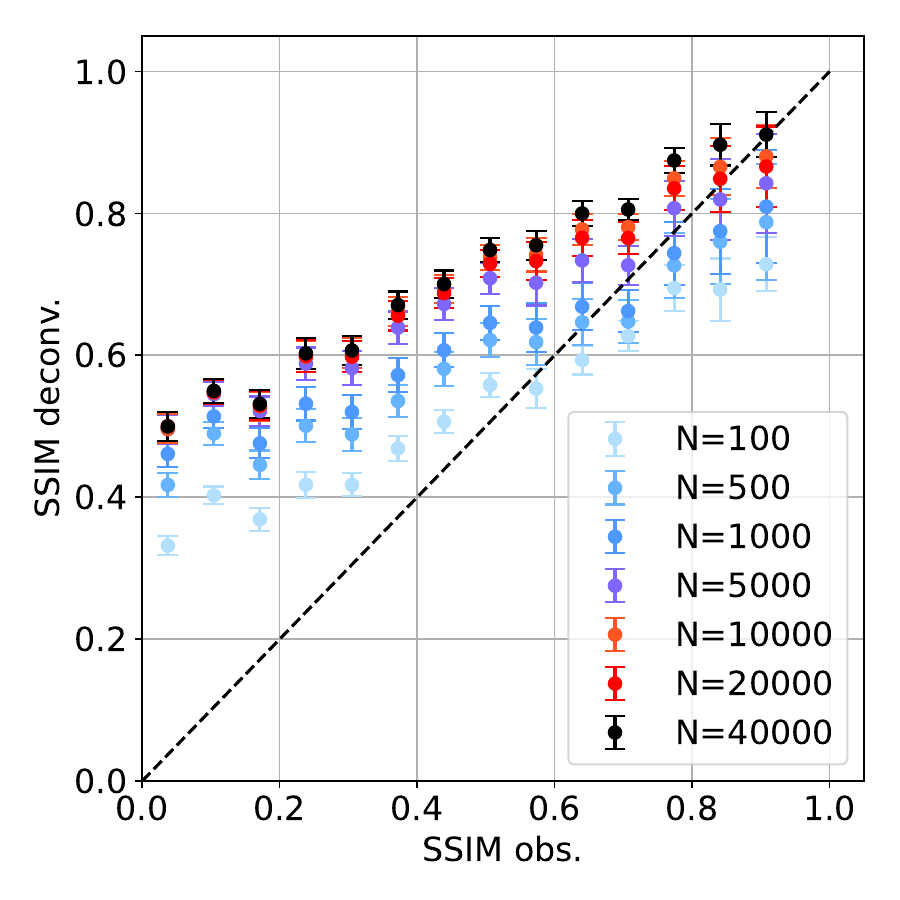}
    \caption{\textbf{Left}: PSNR of deconvolved images as a function of the PSNR of the observed images. The color gradient of the data points, ranging from light blue to black, distinguishes U-Net models trained with increasing sizes of the training dataset ($N=100$, $N=500$, $N=1000$, $N=5000$, $N=10000$, $N=20000$, and $N=40000$). The error bars represent the standard error on the mean in each bin of the profile histogram. \textbf{Right}: same legend but considering the SSIM metric. In each subfigure, the dashed line represents the  case where the PSNR (or SSIM) of the deconvolved images is equal to the PSNR (or SSIM) of the observed images.}
    \label{fig:psnr-ssim_unet_vs_psnr-ssim_obs_ntrain}
\end{figure*}

We now assess the model's performance on a validation set of $1000$ images.
Besides the PSNR metric, we also use the Structural Similarity Index (SSIM)\footnote{We have used the \texttt{structural\_similarity} function of the \texttt{scikit-image} library.}--even if introduced first by \cite{Zhou2004} in the context of natural images--  which provides a number in the range $[0,1]$ comparing the structural information or spatial
interdependencies: the higher is the index value, higher is the similarity between two images. Furthermore, we investigate the impact of varying the number of training images from $100$ up to $40,000$. 

The results are presented in Figure~\ref{fig:psnr-ssim_unet_vs_psnr-ssim_obs_ntrain}. The left subfigure shows the PSNR of the U-Net deconvolved images as a function of the PSNR of the observed images, while the right subfigure illustrates the same relationship for the SSIM metric. In both subfigures, the color gradient of the data points—ranging from light blue to black—represents U-Net models trained with increasing training dataset sizes (see figure caption for specific values). The error bars indicate the standard error on the mean for each bin of the profile histogram.

First, let us consider the case where the U-Net model has been trained with $N=40,000$ images, as discussed in the previous section. The corresponding data points, represented by black dots, are all positioned above the diagonal dashed line. This indicates that the U-Net consistently achieves an improvement in image restoration, particularly for observed images with low PSNR and SSIM values. This trend underscores the model's effectiveness in enhancing image quality even under challenging conditions.

Secondly, the quality of image restoration clearly improves with the number of training images, reaching near-optimal performance for $N > 5,000$. This highlights the need for a sufficiently large training set to fully exploit the model's capabilities, a point further elaborated in the next section through cosine similarity analysis.
\subsection{The Cosine Similarity: Comparison of Two U-Net Models}
\label{sec:cosine_similarity}
\begin{figure*}
    \centering
    \includegraphics[width=0.75\linewidth]{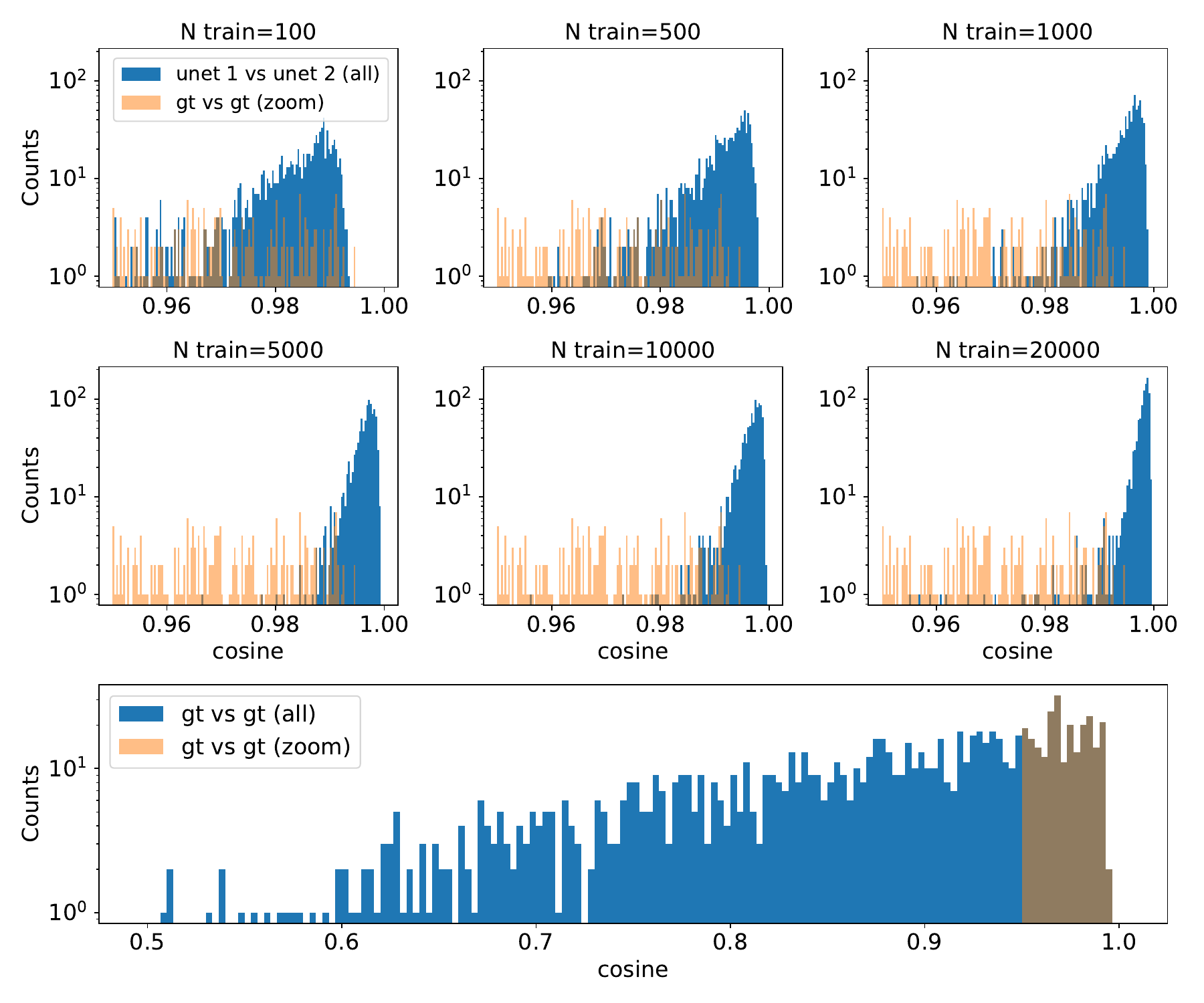}
    \caption{Cosine similarity distributions for two U-Net models trained with varying dataset sizes. The blue histograms represent the cosine similarity between the outputs of \texttt{unet$_1$} and \texttt{unet$_2$}, while the orange histograms show the maximum cosine similarity between two ground truth images.}
    \label{fig:cosine_sim_2Unets_gtgt}
\end{figure*}

The previous section suggests that, with $N > 5,000$ training images, the PSNR and SSIM performances tend to saturate. However, one may question whether this saturation behavior persists when comparing the outputs of two independently trained U-Net models. Specifically, we address the following question: Do two U-Net models, trained with two disjoint training sets of the same size, produce similar restored images from the same observed image? This question is inspired by our previous analysis in the context of galaxy image generation \citep{Campagne_2025}.

To investigate this, we examine the \textit{cosine similarity} between the outputs of two U-Net models, denoted as \texttt{unet$_1$} and \texttt{unet$_2$}. For this purpose, the $40,000$ training images are split into pairs of disjoint subsets (\texttt{S$_1$},  \texttt{S$_2$}) of equal size, ranging from $100$ to $20,000$ images, to independently train \texttt{unet$_1$} and \texttt{unet$_2$}. The results are presented in Figure~\ref{fig:cosine_sim_2Unets_gtgt}.

The first two rows display the cosine similarity distributions (blue histograms) for U-Net models trained with increasing dataset sizes. These blue histograms are derived from the $1,000$ validation images. For comparison, the distributions of the maximum cosine similarity between two ground truth images are also shown (orange histograms, labeled "zoom").

In the last row, we present the distribution of the maximum cosine similarity for \textit{all} ground truth images (blue histogram), with an overlay zooming into the range $[0.95, 1.0]$, as used in the first two rows.

As the training set size increases, a peak in the cosine similarity blue histograms emerges and moves closer to the maximal value (i.e., $1$), indicating that the outputs of the two U-Net models become increasingly similar. Notably, the histogram obtained with only the ground truth images (orange) does not exhibit such a peak, indicating that there are no pairs of highly similar ground truth images. Even beyond $N=5,000$, the cosine similarity continues to increase when considering $N=20,000$. Therefore, we anticipate that with $N=40,000$, the similarity would be even higher, although we cannot test this in practice due to the limited number of available ground truth images. Additionally, this numerical experiment reinforces the extra confidence in U-Net outputs using the two-models framework.
\subsection{U-Net Blind Deconvolution versus Tikhonov Non-Blind Deconvolution}
\label{sec:unet_vs_tikhonov}
As demonstrated in the previous sections, the end-to-end blind deconvolution performed by the U-Net model proves to be highly effective when trained with a sufficient number of images. In Section~\ref{sec:math}, we introduced the Tikhonov functional (Equation~\ref{eq:Tikhonov}) in the context of non-blind deconvolution. Despite its inherent limitations, we employ the Tikhonov approach as an oracle-like comparator, leveraging perfect knowledge of both the true PSF and the ground truth images. This allows us to establish a reference for evaluating the blind deconvolution performance of the U-Net model.

To implement this comparison, we use the Laplacian as the regularization operator and leverage the image generation process described in Equation~\ref{eq:the_generators} and Section~\ref{sec:image_forward_model}. The true PSF is provided\footnote{Note that the computations are performed in the Fourier space.} as input to compute the deconvolved image using Equation~\ref{eq:tikhonov_solution}. The hyperparameter $\lambda$ is optimized to maximize PSNR between the ground truth and the deconvolved images.

\begin{figure*}
    \centering
    \includegraphics[width=0.35\linewidth]{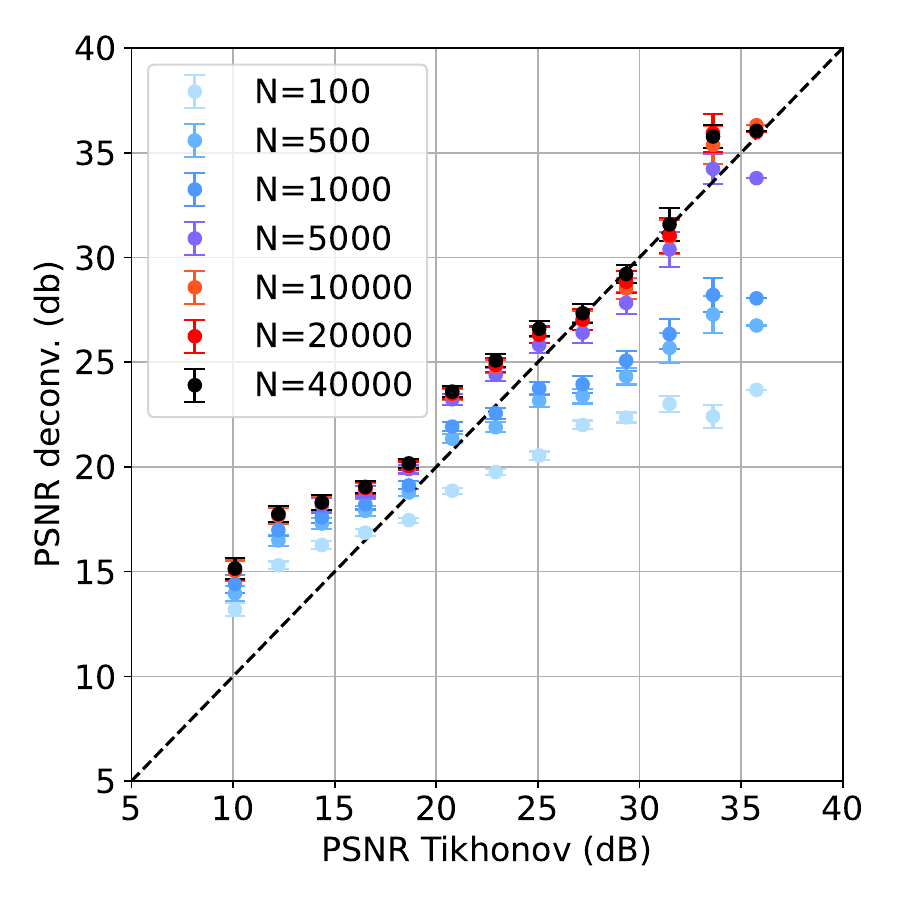}
    \includegraphics[width=0.35\linewidth]{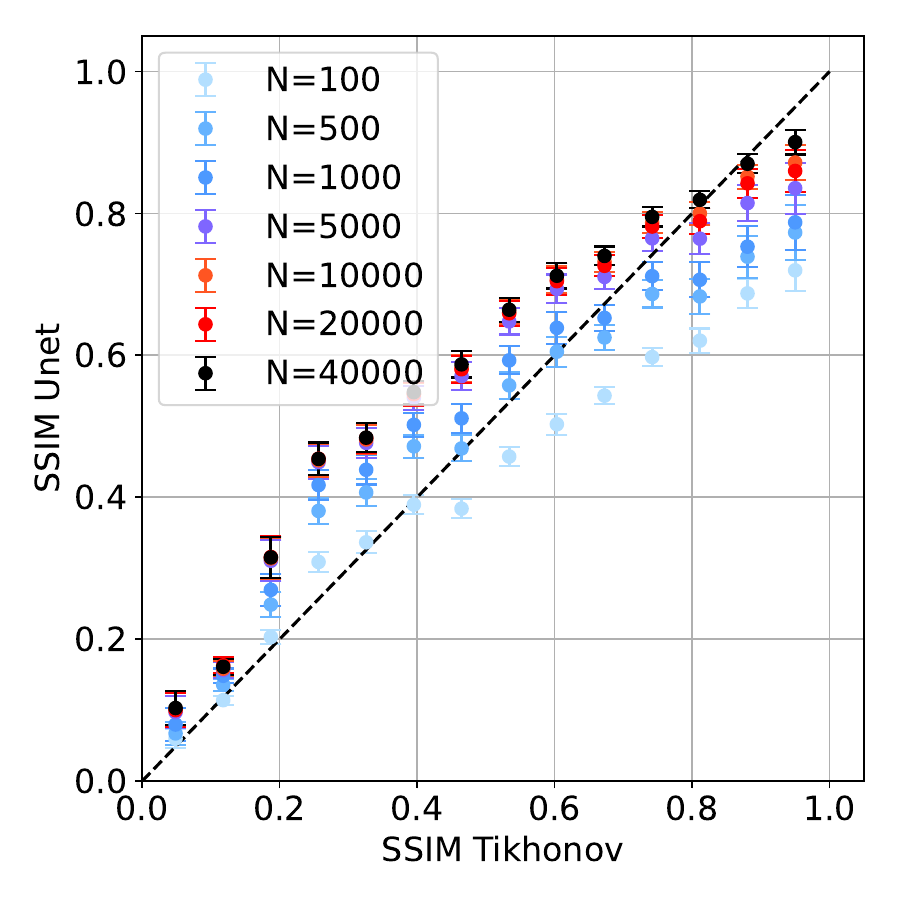}
    \caption{\textbf{Left}: PSNR comparison between U-Net blind deconvolution and Tikhonov non-blind deconvolution for different training set sizes. The horizontal axis represents the PSNR of the Tikhonov output with respect to the ground truth, while the vertical axis represents the PSNR of the U-Net output. The legend indicates the number of training images used for the U-Net model. \textbf{Right}: same legend but considering the SSIM metric. In each subfigure, the dashed line represents the case where the PSNR (or SSIM) of the deconvolved images is equal to the PSNR (or SSIM) of the Tikhonov deconvolved images.}
    \label{fig:psnr-sim_unet_vs_psnr_wiener_ntrain}
\end{figure*}

The results using the U-Net models trained with variable training set sizes, as described in Section~\ref{sec:PSNR_SSIM_vs_Ntrain_perf}, are presented in Figure~\ref{fig:psnr-sim_unet_vs_psnr_wiener_ntrain}. The left subfigure compares the PSNR metric, while the right subfigure compares the SSIM metric, both between the U-Net blind deconvolution and the Tikhonov non-blind deconvolution.

The results confirm that the U-Net blind deconvolution performance improves consistently with an increasing number of training images, in line with the conclusions presented in Sections~\ref{sec:PSNR_SSIM_vs_Ntrain_perf} and~\ref{sec:cosine_similarity}.  Notably, when trained with $40,000$ images (black dots), the U-Net blind deconvolution not only achieves performance levels comparable to the non-blind oracle-like Tikhonov deconvolution but also significantly outperforms it in challenging conditions, specifically at low PSNR and medium SSIM values.  This unexpected outcome underscores the robustness of the U-Net blind deconvolution approach when trained using the methodology described in Sections~\ref{sec:UNet_training} and~\ref{sec:image_forward_model}.
\subsection{Generalization: Seeing and Noise Standard Deviation}
\label{sec:generalisation}
\begin{figure*}
    \centering
    \includegraphics[width=0.35\linewidth]{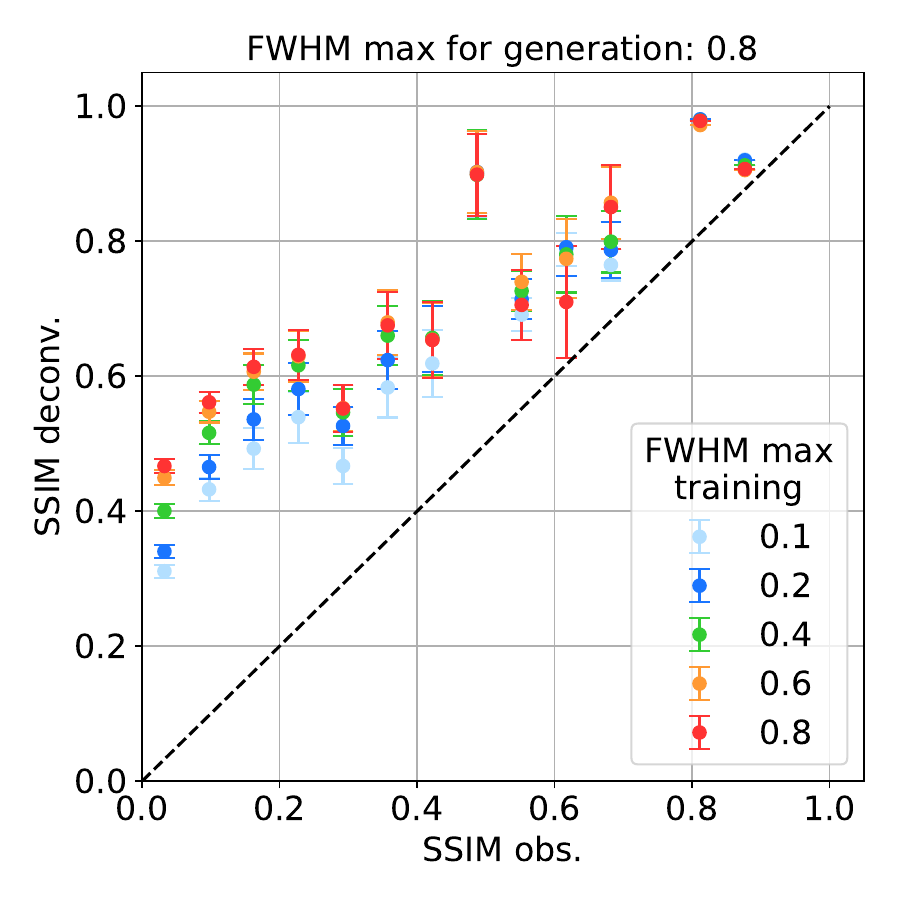}
    \includegraphics[width=0.35\linewidth]{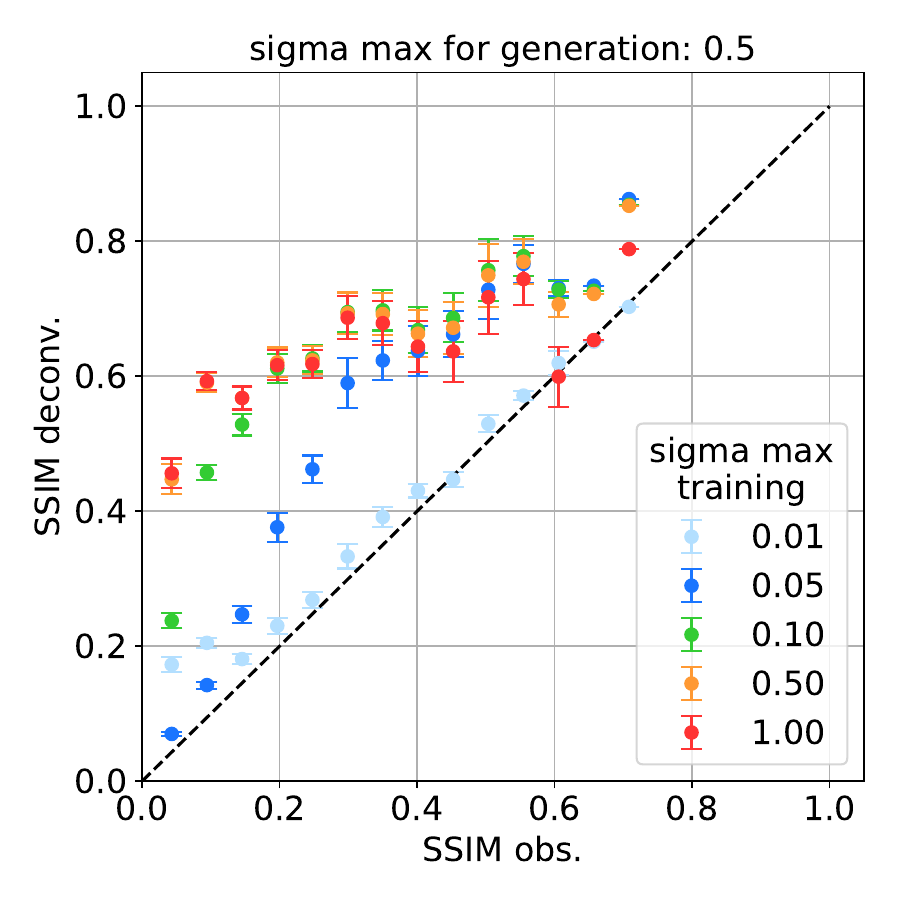}
    \caption{Generalization performance of the U-Net blind deconvolution model. \textbf{Left:} SSIM of deconvolved images as a function of the SSIM of observed images, for validation images generated with a maximum FWHM of 0.8. The color gradient of the data points indicates the maximum FWHM used during training, ranging from 0.1 (light blue) to 0.8 (red). \textbf{Right:} SSIM of deconvolved images as a function of the SSIM of observed images, for validation images generated with a noise standard deviation ($\sigma$) with a maximum value of 0.5 and a minimum value of 0.05. The color gradient indicates the maximum $\sigma$ used during training, ranging from 0.01 (light blue) to 1.0 (red). Error bars indicate the standard error on the mean for each bin.}
    \label{fig:SSIM_Unet_deconv_vs_obs_FWHM_SIGMA_MAX}
\end{figure*}

In the previous sections, all datasets---training, testing, and and validation---were generated (as described in Section~\ref{sec:image_forward_model}) using default maximal values for the Full Width at Half Maximum (FWHM) of the Kolmogorov function (representing the seeing) and the standard deviation of the Gaussian white noise. As a reminder, these values were set to $\texttt{atmos\_fwhm\_max} = 0.4$ and $\texttt{sigma\_noise\_max} = 1.0$, respectively (see Table~\ref{tab:generator_distrib}).

In this section, we investigate the generalization capability of a trained U-Net model under two distinct scenarios:
\begin{itemize}
\item \textit{Varying the atmospheric seeing:}
In this scenario, we evaluate the performance of U-Net models trained with the default $\texttt{sigma\_noise\_max} = 1.0$ and variable $\texttt{atmos\_fwhm\_max}$ values ranging from $0.1$ to $0.8$. The models are then applied to validation images generated with $\texttt{atmos\_fwhm\_max}=0.8$ and the default $\texttt{sigma\_noise\_max}=1.0$.

The left plot of Figure~\ref{fig:SSIM_Unet_deconv_vs_obs_FWHM_SIGMA_MAX} shows that the SSIM of the deconvolved images is relatively stable with respect to the maximum FWHM value used during training. However, models trained with FWHM values closer to those of the validation images (e.g., $\texttt{atmos\_fwhm\_max}=0.8$, shown in red) achieve significantly better performances.

\item \textit{Varying the noise level:}
In this scenario, we evaluate the performance of U-Net models trained with the default $\texttt{atmos\_fwhm\_max}= 0.4$ and variable $\texttt{sigma\_noise\_max}$ ranging from $0.01$ to $1.0$. The models are then applied to validation images generated with the default $\texttt{atmos\_fwhm\_max}= 0.4$ and with a noise standard deviation distributed uniformly \footnote{This is a bit different from Table~\ref{tab:generator_distrib} settings.} in the range $[0.05,0.5]$.

The right plot of Figure~\ref{fig:SSIM_Unet_deconv_vs_obs_FWHM_SIGMA_MAX} reveals that the U-Net model maintains robust performance only if trained with noise levels equal to or higher than those of the validation images (i.e., $\sigma \geq 0.5$). We notice a sharp drop in performance with models trained with $\sigma < 0.5$, especially at low observed SSIM. It is noticeable that for the model trained with $\sigma=0.01$, the outputs have the same SSIM as the observed images, indicating a lack of deconvolution power.
\end{itemize}
These observations highlight both the robustness and the limitations of the U-Net model, which is capable of adapting to varying seeing conditions. Concerning the noise level, the test emphasizes the importance of training with noise conditions representative of the validation conditions.
\section{Summary and Discussion}
\label{sec:summary_discussion}
\subsection{Summary}
In this work, we addressed the \textit{core question} introduced in Section~\ref{sec:unet_deconv_question}: \textit{Can a U-Net architecture, originally designed for image segmentation and denoising, perform end-to-end blind deconvolution of astronomical images without relying on hybrid pipelines or prior knowledge of the Point Spread Function (PSF) and noise characteristics?} To answer this question, we designed a comprehensive experimental framework that simulates realistic astronomical observations using the \texttt{GalSim} toolkit, including random transformations, convolution with PSFs accounting for optical and atmospheric effects, and the addition of Gaussian white noise. We trained a U-Net model using a Mean Square Error (MSE) loss function and evaluated its performance as a function of the training dataset size and its ability to generalize under varying seeing conditions and noise levels. 

Our results, presented in Sections~\ref{sec:deconv_some_illustrations},~\ref{sec:PSNR_SSIM_vs_Ntrain_perf}, and~\ref{sec:cosine_similarity}, demonstrate that the U-Net model is quite effective in deconvolving galaxy images, even under challenging conditions. We observed that the model's performance improves consistently with an increasing number of training images, with PSNR and SSIM metrics tending to saturate for training sets larger than $5,000$ images. The cosine similarity analysis (Section~\ref{sec:cosine_similarity}) further revealed that two independently trained U-Net models produce increasingly similar outputs as the training set size increases, suggesting convergence towards a stable solution. Notably, the similarity continues to increase beyond the $5,000$ training size, indicating that the cosine similarity of the output models is a more sensitive metric, as previously explored in \cite{Campagne_2025}.

In Section~\ref{sec:unet_vs_tikhonov}, we compared the performance of the U-Net model with the oracle-like Tikhonov non-blind deconvolution method, which assumes perfect knowledge of the PSF and ground truth images. Surprisingly, the U-Net model not only matched the performance of the Tikhonov method but also  outperformed it in challenging conditions, particularly at low PSNR and medium SSIM values. This outcome directly addresses our core question by demonstrating that the U-Net architecture can indeed perform end-to-end blind deconvolution effectively.

Finally, in Section~\ref{sec:generalisation}, we assessed the generalization capability of the U-Net model under varying seeing conditions and noise levels. Our results reveal that the model demonstrates a notable ability to adapt to unseen seeing conditions (FWHM), maintaining consistent performance even when the validation FWHM differs from the training parameters. However, its robustness to noise variations is more constrained: the model performs optimally only when trained with noise levels equal to or higher than those encountered during validation. This highlights both the flexibility of the U-Net architecture in handling atmospheric distortions and the critical importance of matching noise conditions between training and application scenarios.
\subsection{Discussion}
\label{sec:Discussion}
While hybrid methods combining traditional and deep learning approaches have shown success in previous studies, such as in \cite{Sureau2020, Akhaury2024}, this work focuses on exploring the intrinsic capabilities of a \textit{standalone} U-Net model for end-to-end blind deconvolution in astronomical imaging. Although using a U-Net alone may not represent the current state-of-the-art, our results highlight its remarkable potential in this context. The ability of the U-Net model to generalize across varying conditions of seeing and noise, and to outperform the oracle-like Tikhonov deconvolution in certain challenging scenarios, provides a positive answer to our core question.

The findings of \cite{kadkhodaie2024generalization} provide a valuable framework for interpreting our results. The authors characterized the U-Net network as learning a form of \textit{geometry-adaptive harmonic basis} during denoising tasks, extending beyond traditional steerable wavelet bases. Although a rigorous mathematical analysis of the U-Net's behavior in end-to-end blind deconvolution is not yet available to our knowledge, our experiments with $C^\alpha$ images (Appendix~\ref{sec:calpha_deconv}) suggest a similar mechanism. The eigenvectors of the U-Net's Jacobian exhibit oscillating patterns, adapting to the geometry of the input image, and the decomposition in the eigenbasis is sparse. In classical non-blind deconvolution (Section~\ref{sec:math}), sparse representations, particularly through multi-resolution wavelet analysis, play a crucial role. If the U-Net is indeed learning such adaptive sparse representations during deconvolution, it could explain its ability to effectively reveal the ground truth image, warranting further theoretical investigation.

However, some caveats must be acknowledged. While the U-Net model demonstrates valable  performance, its effectiveness is still influenced by the similarity between training and validation conditions. Although we employed a probabilistic forward model (Sections~\ref{sec:UNet_training}, \ref{sec:image_forward_model}), the deconvolution performed by the U-Net is not entirely blind, as the training data implicitly encodes certain characteristics of the PSF and noise distributions. This suggests that the network's ability to generalize is contingent on the representativeness of the training data.

Future work could extend this study by employing larger datasets composed of larger and more diverse and structurally complex galaxy images. Additionally, exploring the transferability of these techniques to other domains could provide further insights into what the network learns during the image deconvolution process. Such investigations may help confirm whether the U-Net's performance is domain-specific or indicative of a more general capability to learn adaptive sparse representations as suggested by our experiments with $C^\alpha$ images. Another avenue would be to replace the classical U-Net architecture, which relies on convolutional kernels, with more advanced variations such as those proposed in \cite{Akhaury2024}. 

Finally, notice that, unlike the objective in Equation~\ref{eq:blind}, which aims to jointly estimate the Point Spread Function (PSF) and the deconvolved image, our current U-Net architecture only outputs the deconvolved image, without providing explicit information about the blurring kernel. This limitation opens a potential research avenue to investigate whether the U-Net's learned sparse representation could implicitly encode information about the PSF.

In conclusion, this work demonstrates the potential of U-Net architectures for end-to-end blind deconvolution in galaxy imaging, particularly for small image patches. It lays the groundwork for further research into the underlying mechanisms of the U-Net's deconvolution capabilities and opens avenues for exploring its applicability across different imaging domains.
\section{Reproducible Research}
\label{sec:data_availability}
The companion repository on GitHub (\url{https://github.com/jecampagne/Blind-deconvolution-compagnon}) contains all the codes for network training, data preprocessing, and plot generation. The COSMOS Real Galaxy Dataset, used for training and validation, is available on Zenodo (\url{https://zenodo.org/records/3242143}).
The code is released under the MIT License, and the COSMOS dataset is available under a CC-BY 4.0 license.
\section*{Large Language Model Use Disclosure}
We used Le Chat CNRS Enterprise, an AI-powered language model developed by Mistral AI\footnote{\url{https://mistral.ai}} and tailored to the research needs of the National Centre for Scientific Research (CNRS), to assist in refining the clarity and fluency of the English text. All AI-generated suggestions were carefully reviewed and validated by the author.
\section*{Acknowledgments}
We  thank the French GENCI--IDRIS for providing access to Nvidia V100 resources under Grant 2025-AD010413957R2.
Our code for processing galaxy images is inspired by the work of \cite{Li2023mnras} and is available at \url{https://github.com/Lukeli0425/Galaxy-Deconv/} under the MIT License. The code for handling $C^\alpha$ images is inspired by the work of \cite{kadkhodaie2024generalization} and is available at \url{https://github.com/LabForComputationalVision/memorization_generalization_in_diffusion_models}  under the MIT License.
\bibliographystyle{apsrev4-1}
\bibliography{refs}
%
\begin{appendix}
\section{Generator parameters}
\label{sec:appendix_random_distrib}
In Sections~\ref{sec:UNet_training}, \ref{sec:image_forward_model} the images provided for training, testing and validation phases are based on random distributions. The Table~\ref{tab:generator_distrib} gives the details.  
\begin{table}[h]
    \renewcommand{\arraystretch}{1.5}
    \centering
    \caption{Random number distributions with their default values used to generate training, testing, and validation datasets. $\mathcal{U}(a,b)$ denotes the uniform distribution over the interval $[a,b] \subset \mathbb{R}$, while $\mathcal{U}(\{a,b,\dots\})$ is the discrete uniform distribution over the finite set $\{a,b,\dots\}$. $\mathcal{N}(\mu,\sigma)$ represents the normal distribution with mean $\mu$ and standard deviation $\sigma$.}
    \label{tab:generator_distrib}
    \begin{tabular}{p{0.3\linewidth}cc}
    \textbf{Parameter} & \textbf{Distribution} & \textbf{Default value}\\ \hline\hline
      \centering \textit{Atmospheric PSF} & & \\
      FWHM of the Kolmogorov function (arcsec) & $\texttt{atmos\_fwhm\_max} \times \mathcal{U}(0,1)$& $0.4$\\
      Ellipticity magnitude of the shear & $\texttt{atmos\_e}\times \mathcal{U}(1,2)$ & $0.01$\\
      Shear position angle (radians) & $2\pi \times \mathcal{N}(0,1)$ & \\
      \centering\textit{ Optical PSF} & & \\
      Defocusing  & $\mathcal{N}(0,\texttt{sigma\_defocus})$ & $ 0.10$ \\
      Astigmatism (axes $x$ \& $y$) & $\mathcal{N}(0,\texttt{sigma\_opt\_psf})$ & $ 0.07$\\
      Coma (axes $x$ \& $y$) & $\mathcal{N}(0,\texttt{sigma\_opt\_psf})$ & " \\
      Spherical aberration & $\mathcal{N}(0,\texttt{sigma\_opt\_psf})$ & " \\
      Trefoil (axes $x$ \& $y$) & $\mathcal{N}(0,\texttt{sigma\_opt\_psf})$ & "  \\
      Obscuration & $\texttt{opt\_obs\_min} + \texttt{opt\_obs\_width}\times \mathcal{U}(0,1)$ & $(0.1, 0.4)$\\
      $\lambda/D$ & $\texttt{lam\_ov\_d\_min} + \texttt{lam\_ov\_d\_width}\times \mathcal{U}(0,1)$ & $(0.017,0.007)$\\
      \centering \textit{Image galaxy parameters} & & \\
      Magnitude of the shear & $pdf(x)=x;\ s.t.\ x \in [\texttt{min\_shear,max\_shear}]$ & $(0.01,0.05)$\\
      Shear position angle (radians) &  $2\pi \times \mathcal{N}(0,1)$\\
      Magnification & $\mathcal{U}(1,1.1)$\\
      Rotation angle (radians) & $\mathcal{U}(\{0,\pi/2,\pi,3\pi/2\})$\\
      Offset along $x$ and $y$ axes (pixel unit) & $\mathcal{U}(-1,1)$\\
      Standard deviation of the Gaussian white noise & $\texttt{sigma\_noise\_max}\times \mathcal{U}(0,1)$ &
      1.0\\
      \hline\hline
    \end{tabular}
\end{table}
\section{Extra optical PSF examples}
\label{sec:aap_optical_PSF}
\begin{figure}[h]
    \centering
    \includegraphics[width=0.6\linewidth]{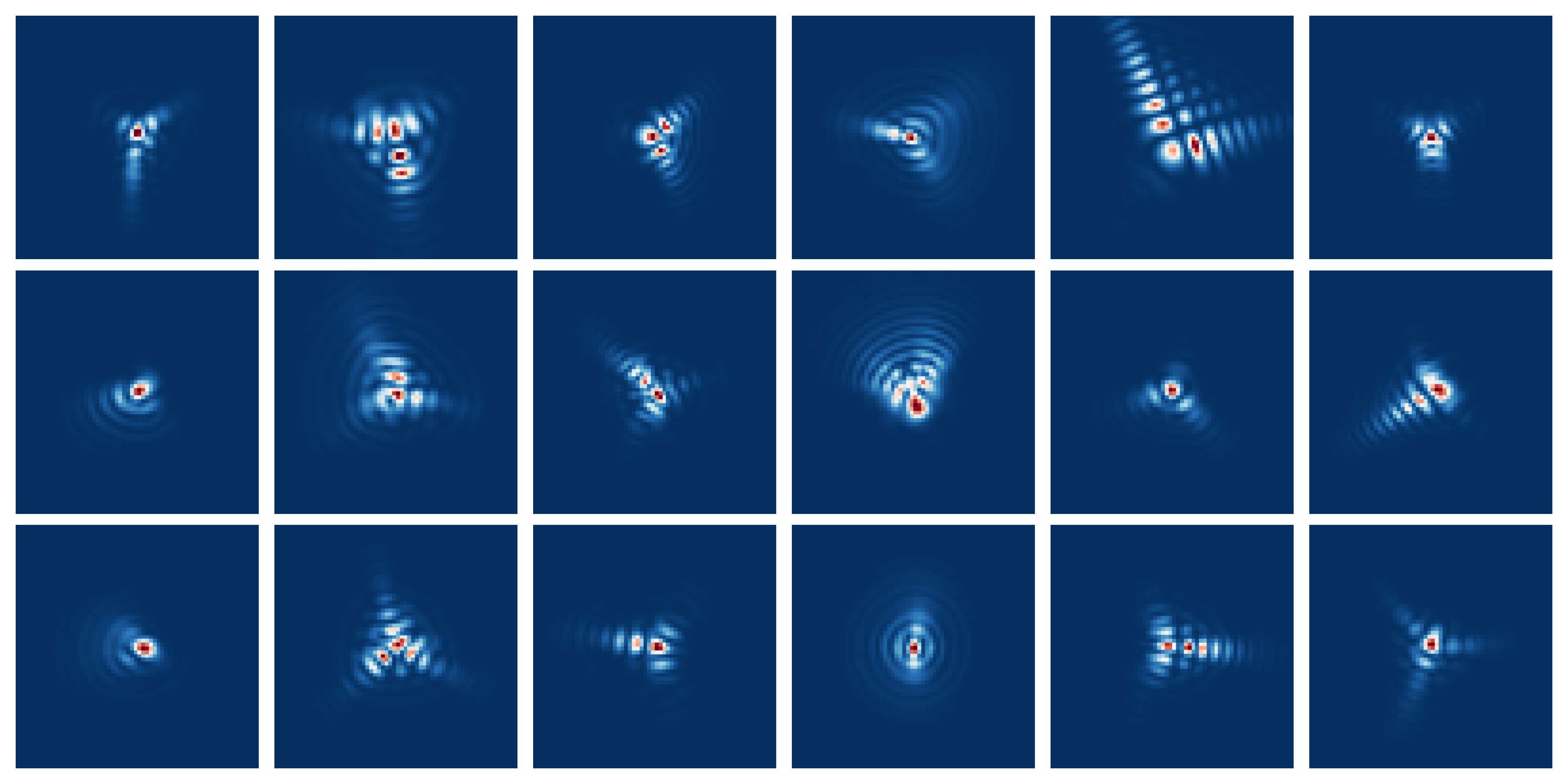}
    \caption{Example of optical PSFs (zoom $\times 3$) obtained from \texttt{galsim.OpticalPSF} (see Section~\ref{sec:image_forward_model} for details).}
    \label{fig:psf_optics_exemple_18}
\end{figure}

The Figure~\ref{fig:psf_optics_exemple_18} presents some examples of Point Spread Function generated with some telescope optical characteristics complementing the example shown in Section~\ref{sec:image_forward_model}.
\section{U-Net blind deconvolution with $C^\alpha$ synthetic images}
\label{sec:calpha_deconv}
\begin{figure}
    \centering
    \includegraphics[width=0.6\linewidth]{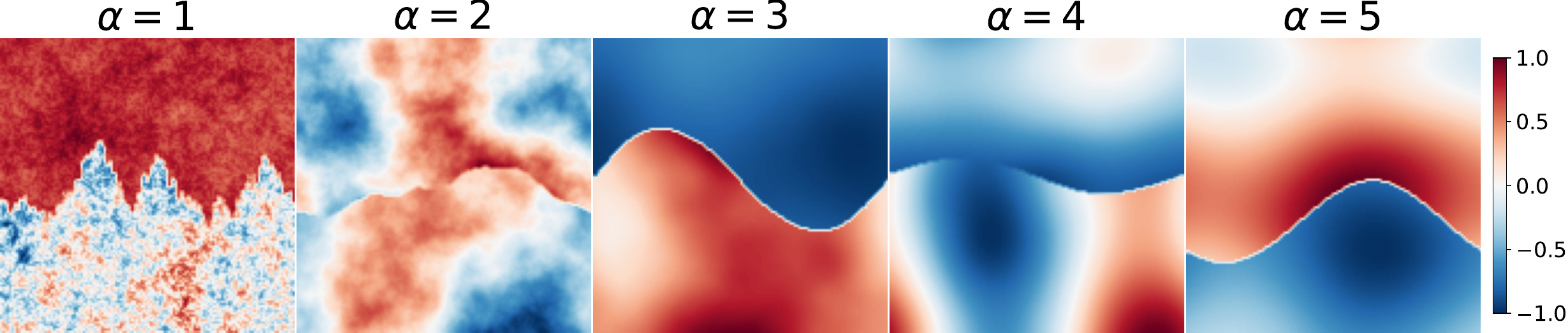}
    \caption{Examples of $C^\alpha$ images with varying regularity. Note that within a single image, the regularity of the backgrounds matches that of the frontier.}
    \label{fig:calpha_examples}
    \vspace{1em}
\end{figure}

The question raised in the Section~\ref{sec:Discussion} on the possible geometrical interpretation of what is learned by the U-Net may be addressed following the work by  \cite{kadkhodaie2024generalization}, considering geometric $C^\alpha$ images. These synthetic images consist of a regular frontier between regular backgrounds, where the degree of regularity is controlled by the parameter $\alpha$, see examples in Figure~\ref{fig:calpha_examples}.

We conducted the following experiment. First, we generated datasets of more than $100,000$ images of $d=128 \times 128$ pixels, with pixel values adjusted to the range $[-1,1]$ and with regularity ranging from $\alpha=1$ to $\alpha=5$. Then, for each $\alpha$-dataset, we employed a U-Net architecture with an increased number of blocks (i.e., $4$), which differs from the default configuration used in our main experiments, to better handle the larger image size according to Table~1 of \cite{kadkhodaie2024generalization}. Unlike \cite{kadkhodaie2024generalization}, we use architectures with biases in both the BatchNorm and Convolutional layers\footnote{The authors motivate the use of a bias-free architecture to simplify interpretation but argue that the learned sparse representation results remain unaffected by the inclusion of biases.}.

\begin{figure}[h]
    \centering
    \includegraphics[width=0.6\linewidth]{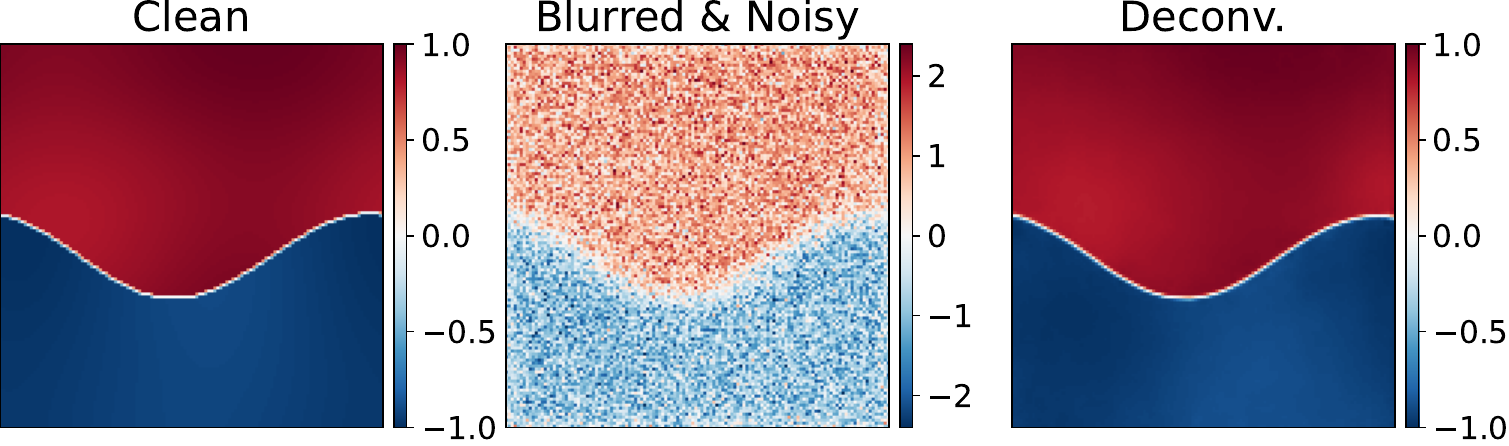}
    \caption{\textbf{Left}: A clean $C^5$ image, \textbf{Middle}: The image convoluted with a Gaussian kernel of $\texttt{FWHM}=10$ to which is added a Gaussian white noie of $\sigma=0.5$. \textbf{Right}: The image blindly deconvoluted.}
    \label{fig:calpha_blur_noisy_deconv_imgs}
\end{figure}

 The models were trained using the same methodology as described in Section~\ref{sec:UNet_training}, employing forward modelling. However, in the context of the synthetic images considered here, we applied convolution using Gaussian blur kernels with random Full Width at Half Maximum (FWHM) in the range $[0,10]$ pixels. Note that kernels with FWHM values below $1$ pixel are replaced by an identity kernel. The i.i.d. additive Gaussian noise had a standard deviation $\sigma$ randomly sampled from the range $[0,0.5]$. An example of the deconvolution of a $C^5$ image is given in Figure~\ref{fig:calpha_blur_noisy_deconv_imgs}.

\begin{figure}[h]
    \centering
    \includegraphics[width=0.4\linewidth]{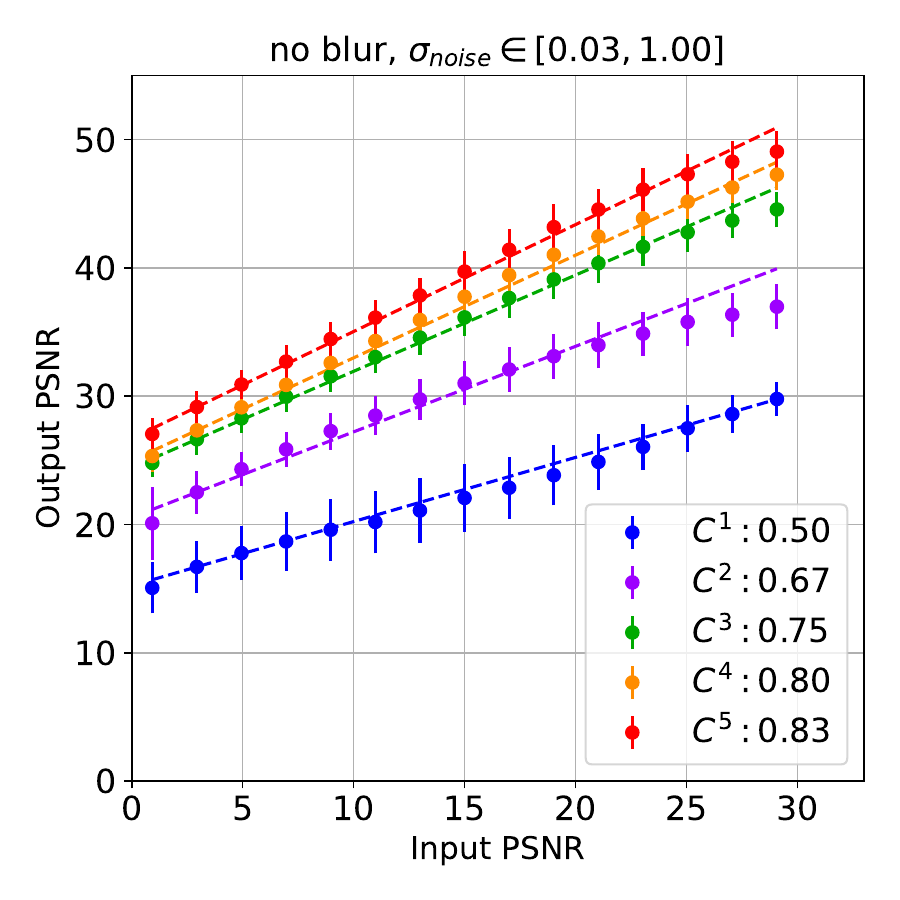}
    \caption{Inspired by Figure~4 of \cite{kadkhodaie2024generalization}, this figure shows results for a U-Nets trained for end-to-end blind deconvolution. PSNR of the U-Net output as a function of the input PSNR, for U-Nets tested on images corrupted solely by additive white Gaussian noise with varying $\sigma$ (i.e. no blurring). The dashed lines represent the theoretical asymptotic slopes of the optimal denoiser, $\alpha/(1+\alpha)$.}
    \label{fig:calpha_psnr_unet_vs_psnr_obs_noblur}
\end{figure}
One of the key results in \cite{kadkhodaie2024generalization}, where U-Net is used solely as a denoiser, is that the trained U-Net reaches the optimal denoiser PSNR, which asymptotically increases with a slope of $\alpha/(1+\alpha)$ as a function of the input PSNR. Along these lines, Figure~\ref{fig:calpha_psnr_unet_vs_psnr_obs_noblur} shows the deconvolved PSNR as a function of the input PSNR for U-Nets trained blindly as described above and tested on images that were not blurred but only corrupted by additive white Gaussian noise with $\sigma$ such that the PSNR ranges from $0$ to $30$. We observe that the blind-deconvolving U-Net performs as well as the optimal denoiser in a pure denoising use case. In the context of a deconvolution task, to our knowledge there is no equivalent of the asymptotically slope relation. 

Another key result in \cite{kadkhodaie2024generalization} concerns the eigendecomposition of the Jacobian of the denoiser. We briefly recall the key ideas. In the context of a bias-free architecture, one can write (for notational simplicity, $\bm{y}$ is the observed noisy image, equivalent to $\bm{x}_o$ used in Equation~\ref{eq:the_pb}):
\begin{equation}
    \bm{\hat{x}}(\bm{y}) = \mathcal{D}(\bm{y}) = \nabla_{\bm{y}} \mathcal{D}(\bm{y}) \ \bm{y}.
\end{equation}
Let $(\lambda_k(\bm{y}), \bm{e}_k(\bm{y}))_{1 \leq k \leq d}$ denote the eigenvalues and eigenvectors of $\nabla_{\bm{y}} \mathcal{D}(\bm{y})$, which depend on the observed corrupted image $\bm{y}$. Projecting $\bm{y}$ onto the eigenvector basis leads to
\begin{equation}
    \bm{\hat{x}}(\bm{y}) = \sum_k \lambda_k(\bm{y}) \ \langle \bm{y}, \bm{e}_k(\bm{y}) \rangle \ \bm{e}_k(\bm{y}),
    \label{eq:decomposition_ekbasis}
\end{equation}
where $\langle \cdot, \cdot \rangle$ denotes the inner product. The denoiser can then be interpreted as performing shrinkage in an adaptive basis, with $\lambda_k(\bm{y})$ acting as a threshold, similar to denoising in a fixed basis \cite{mallat2009wavelet}. Even if the U-Net we have used is not bias-free, we can apply the eigendecompostion as explained by  \cite{kadkhodaie2024generalization}.

In the context of the denoising-only task, \cite{kadkhodaie2024generalization} shows examples of eigenvectors $\bm{e}_k(\bm{y})$ and observes that the eigenvectors exhibit oscillating patterns both along the contours and in uniformly regular regions, thus adapting to the geometry of the input image. The authors refer to this as a \textit{geometry-adaptive harmonic basis}. Moreover, the decomposition of $\bm{y}$ in the eigenbasis is sparse, with only a few $\langle \bm{y}, \bm{e}_k(\bm{y}) \rangle$ having significant values, and the eigenvalues $\lambda_k(\bm{y})$ decay rapidly, exploiting this sparsity. However, in the context of the deconvolution task, the validity of these properties is questionable.

\begin{figure}[t]
    \centering
    \includegraphics[width=0.8\linewidth]{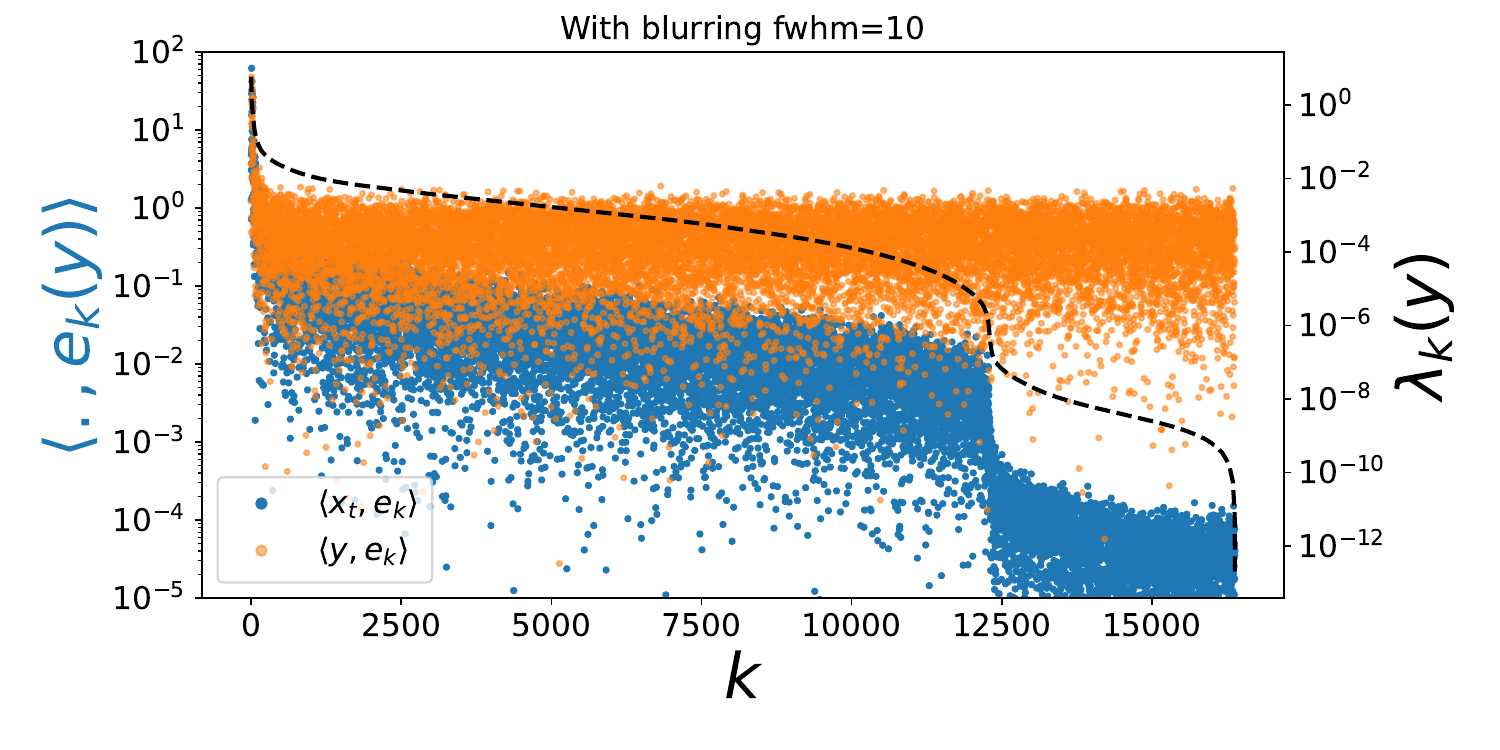}\\
    \includegraphics[width=0.7\linewidth]{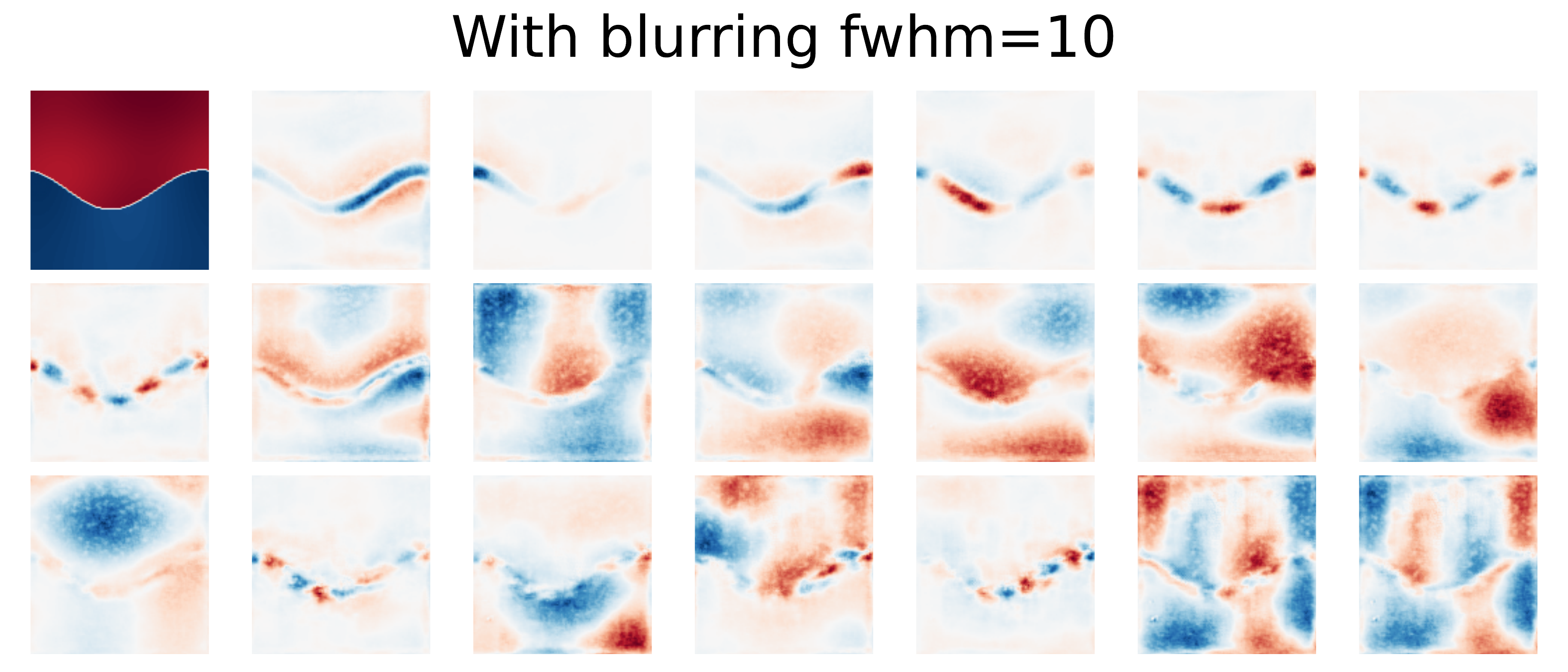}
    \caption{Inspired by Figures~3 and 4  of \cite{kadkhodaie2024generalization}, this figure shows results for a U-Nets trained for end-to-end blind deconvolution.
    \textbf{Top}: Evolution of the $\langle \bm{x}_t, \bm{e}_k(\bm{y}) \rangle$ and  $\langle \bm{y}, \bm{e}_k(\bm{y}) \rangle$ coefficients as well as $\lambda_k(\bm{y})$ as a function of $k$ where $\bm{x}_t$ is a $C^5$-image and   $\bm{y}$ the same image corrupted by a Gaussian kernel with a Full Width at Half Maximum (FWHM) of 10 pixels and additive white Gaussian noise with $\sigma=0.5$. \textbf{Bottom}: The first 20 eigenvectors $\bm{e}_k(\bm{y})$.}
    \label{fig:calpha_sparse_fwhm10}
\end{figure}
Figure~\ref{fig:calpha_sparse_fwhm10} shows, in the top panel, the evolution of the scalar products $\langle \bm{y}, \bm{e}_k(\bm{y}) \rangle$ (Equation~\ref{eq:decomposition_ekbasis}) and $\langle \bm{x}_t, \bm{e}_k(\bm{y}) \rangle$, where $\bm{x}_t$ is the true image, as well as the eigenvalues $\lambda_k(\bm{y})$, all as a function of $k$. The bottom panel displays the first 20 eigenvectors. We observe that both key features---sparsity, with only a few significant scalar products, and the eigenvectors forming a geometry-adaptive harmonic basis---are present when the U-Net is tested on a $C^5$-image corrupted by a Gaussian kernel with a Full Width at Half Maximum (FWHM) of 10 pixels and additive white Gaussian noise with $\sigma=0.5$. Notably, along the frontier the oscillatory patterns exhibit a width consistent with the blur kernel's FWHM. These results have informed the discussion in Section~\ref{sec:summary_discussion}.
\newpage
\end{appendix}
\end{document}